\documentclass[aps,superscriptaddress,twocolumn]{revtex4}
\usepackage[colorlinks, linkcolor=red,anchorcolor=green,citecolor=blue]{hyperref}
\usepackage{graphicx}
\usepackage{bm}
\usepackage{amsmath}
\usepackage{color}
\usepackage{xcolor, color, framed, ulem}
\graphicspath{{./fig/}}

\begin{document}

\newcommand{\com}[1]{{\sf\color[rgb]{1,0,0}{#1}}}

\def\bm{\boldsymbol}

\def\dl{\displaystyle}
\def\du{\end{document}}
\def\d{{\rm d}}
\def\e{{\rm e}}
\def\i{{\rm i}}

\def\esym{$E_{sym}(\rho)$}

\def\amev{MeV/u}
\def\agev{GeV/u}
\def\pacab{${\rm PPAC 1\times 2}$}
\def\pacac{${\rm PPAC 1\times 3}$}

\def\snsn{$^{124}$Sn+$^{124}$Sn}

\title{Properties of the fast fission  and the coincident emissions of light charged particles in  $^{40}$Ar +  $^{197}$Au reactions at 30 MeV/u}
\author{Xinyue Diao}
\affiliation{Department of Physics, Tsinghua University, Beijing 100084, China}
\author{Yijie Wang}
\affiliation{Department of Physics, Tsinghua University, Beijing 100084, China}
\author{Fenhai Guan}
\affiliation{Department of Physics, Tsinghua University, Beijing 100084, China}
\author{Dawei Si}
\affiliation{Department of Physics, Tsinghua University, Beijing 100084, China}
\author{Qianghua Wu}
\affiliation{Department of Physics, Tsinghua University, Beijing 100084, China}
\author{Yan Huang}
\affiliation{Department of Physics, Tsinghua University, Beijing 100084, China}
\author{Liming Lyu}
\affiliation{Department of Physics, Tsinghua University, Beijing 100084, China}
\author{Yuhao Qin}
\affiliation{Department of Physics, Tsinghua University, Beijing 100084, China}
\author{Zhi Qin}
\affiliation{Department of Physics, Tsinghua University, Beijing 100084, China}
\author{Dong Guo}
\affiliation{Department of Physics, Tsinghua University, Beijing 100084, China}
\author{Yaopeng Zhang}
\affiliation{Department of Physics, Tsinghua University, Beijing 100084, China}
\author{Xuan Zhao}
\affiliation{Department of Physics, Tsinghua University, Beijing 100084, China}
\author{Zhen Bai}
\affiliation{Institute of Modern Physics, Chinese Academy of Sciences, Lanzhou 730000, China}
\author{Fangfang Duan}
\affiliation{Institute of Modern Physics, Chinese Academy of Sciences, Lanzhou 730000, China}
\author{\\Limin Duan}
\affiliation{Institute of Modern Physics, Chinese Academy of Sciences, Lanzhou 730000, China}
\author{Zhihao Gao}
\affiliation{Institute of Modern Physics, Chinese Academy of Sciences, Lanzhou 730000, China}
\affiliation{University of Chinese Academy of Sciences, Beijing 100049, China}
\author{Qiang Hu}
\affiliation{Institute of Modern Physics, Chinese Academy of Sciences, Lanzhou 730000, China}
\author{Rongjiang Hu}
\affiliation{Institute of Modern Physics, Chinese Academy of Sciences, Lanzhou 730000, China}
\author{Genming Jin}
\affiliation{Institute of Modern Physics, Chinese Academy of Sciences, Lanzhou 730000, China}
\author{Shuya Jin}
\affiliation{Institute of Modern Physics, Chinese Academy of Sciences, Lanzhou 730000, China}
\affiliation{University of Chinese Academy of Sciences, Beijing 100049, China}
\author{Junbing Ma}
\affiliation{Institute of Modern Physics, Chinese Academy of Sciences, Lanzhou 730000, China}
\affiliation{University of Chinese Academy of Sciences, Beijing 100049, China}
\author{\\Peng Ma}
\affiliation{Institute of Modern Physics, Chinese Academy of Sciences, Lanzhou 730000, China}
\author{Jiansong Wang}
\affiliation{School of Science, Huzhou University, Huzhou, 313000, China}
\affiliation{Institute of Modern Physics, Chinese Academy of Sciences, Lanzhou 730000, China}
\author{Peng Wang}
\affiliation{Institute of Modern Physics, Chinese Academy of Sciences, Lanzhou 730000, China}
\affiliation{University of Chinese Academy of Sciences, Beijing 100049, China}
\author{Yufeng Wang}
\affiliation{Institute of Modern Physics, Chinese Academy of Sciences, Lanzhou 730000, China}
\affiliation{University of Chinese Academy of Sciences, Beijing 100049, China}
\author{Xianglun Wei}
\affiliation{Institute of Modern Physics, Chinese Academy of Sciences, Lanzhou 730000, China}
\author{Herun Yang}
\affiliation{Institute of Modern Physics, Chinese Academy of Sciences, Lanzhou 730000, China}
\author{Yanyun Yang}
\affiliation{Institute of Modern Physics, Chinese Academy of Sciences, Lanzhou 730000, China}
\author{Gongming Yu}
\affiliation{Institute of Modern Physics, Chinese Academy of Sciences, Lanzhou 730000, China}
\affiliation{University of Chinese Academy of Sciences, Beijing 100049, China}
\author{Yuechao Yu}
\affiliation{Institute of Modern Physics, Chinese Academy of Sciences, Lanzhou 730000, China}
\affiliation{University of Chinese Academy of Sciences, Beijing 100049, China}
\author{Yapeng Zhang}
\affiliation{Institute of Modern Physics, Chinese Academy of Sciences, Lanzhou 730000, China}
\author{Qingwu Zhou}
\affiliation{Institute of Modern Physics, Chinese Academy of Sciences, Lanzhou 730000, China}
\affiliation{University of Chinese Academy of Sciences, Beijing 100049, China}
\author{Yaofeng Zhang}
\affiliation{College of Nuclear Science and Technology, Beijing Normal University, Beijing 100875, China}
\author{Chunwang Ma}
\affiliation{Institute of Particle and Nuclear Physics, Henan Normal University, Xinxiang 453007, China}
\author{Xinrong Hu}
\affiliation{University of Chinese Academy of Sciences, Beijing 100049, China}
\affiliation{Shanghai Institute of Applied Physics, Chinese Academy of Science, Shanghai 201800, China}
\author{Hongwei Wang}
\affiliation{Shanghai Institute of Applied Physics, Chinese Academy of Science, Shanghai 201800, China}
\affiliation{Shanghai Advanced Research Institute, Chinese Academy of Science, Shanghai 201210, China}
\author{Artur Dobrowolski}
\affiliation{Uniwersytet Marii Curie Sk{\l}odowskiej, Katedra Fizyki Teoretycznej, Lublin 20031, Poland}
\author{Krzysztof Pomorski}
\affiliation{Uniwersytet Marii Curie Sk{\l}odowskiej, Katedra Fizyki Teoretycznej, Lublin 20031, Poland}
\author{Zhigang Xiao}
\affiliation{Department of Physics, Tsinghua University, Beijing 100084, China}

\date{\today}%

\begin{abstract}
The experiment of Ar+Au reactions at 30 MeV/u have been performed using the Compact Spectrometer for Heavy IoN Experiments (CSHINE) in phase I. The light-charged particles are measured by the silicon stripe telescopes in coincidence with the fission fragments recorded by the parallel plate avalanche counters. The distribution properties of the azimuth difference $\Delta \phi$ and the time-of-flight difference $\Delta TOF$ of the fission fragments are presented varying the folding angles which represents the linear momentum transfer from the projectile to the reaction system. The relative abundance of the light charged particles in the fission events to the inclusive events is compared as a function of the laboratory angle $\theta_{\rm lab}$ ranging from $18^\circ$ to $60^\circ$ in various folding angle windows. The angular evolution of the yield ratios of p/d and t/d in coincidence with fission fragments is investigated. In a relative comparison,  tritons are more abundantly emitted at small angles, while protons are more abundant at large angles. The angular evolution of the neutron richness of the light-charged particles is consistent with the results obtained in previous inclusive experiments.  
\end{abstract}


\maketitle

\section{Introduction}

Nuclear fission is a large-amplitude collective motion mode involving up to hundreds of nucleons. Recently the studies on nuclear fission have been revived for its significance in both nuclear physics and astrophysics. For instance, the competition brought by fission reduces the possibility of the formation of superheavy elements \cite{Guol2018}. In the stellar environment, the abundance of the nuclides in $A\approx 160$ region is significantly influenced by the recycling of the fission products \cite{Lorusso2015,Nishimura2012,Suzuki2012}. Theoretically, statistical fission has been described well by microscopical theories and various phenomenological approaches \cite{PBR96,SCH99,JB2007,LB2011,Zhanghf2014,Tanimura2017,TZL17,WY2018,WY2018-2}. Taking a few recent studies, for example, by computing the potential energy surface in the macroscopic-microscopic model with the Lublin-Strasbourg Drop (LSD) for the macroscopic part and the Yukawa-folded potential for the microscopic energy corrections, the fission fragment mass distributions are predicted in a broad mass region ranging from Thorium to the most superheavy elements, showing rather good consistency with the data if available  \cite{Pomorski2021, Pavel2021}.  Using the time-dependent Hartree-Fock approach, the reactions $^{238}$U+$^{40}$Ca near the Coulomb barrier have been investigated in terms of the influences of various ingredients in effective nuclear interaction, including incompressibility, symmetry energy, effective nucleon mass, etc\cite{Zhengh2018}. 

When the excitation energy or the angular momentum becomes much high, as achieved in heavy-ion reactions well above the Coulomb barrier, the fission barrier tends to vanish. As a consequence,  the fission life-time becomes shorter by a factor of 10 to 100 and the variance of the mass asymmetry increases significantly (the mass asymmetry $\eta$ of the two fragments can be larger than 0.6), compared to the statistical fission \cite{Greg82, Greg82t, Gla83, Leray84, Zheng84}. In addition, the friction grows with the increasing of nuclear temperature and the diffusion enhance the fission probability \cite{Ivanyuk1996, Usang2016,Pomorski2015}. So the process of fast fission is more complicated and the dynamic feature of the fission process is of significance. Usually the transport models have been successfully applied to describe the mass and total kinetic energy distributions and the time scale of the fission \cite{Wen13, Russ11,Riz11,TL09,TO11,LTQ13,TW08,LTO13,WT11,Wuqh2019}. Time-dependent Hartree-Fock theory can give an excellent qualitative and quantitative description of fast fission starting with large deformation\cite{God2015}.

With the advent of advanced $4\pi$ detectors for heavy-ion reactions at Fermi energies, fast fission, usually termed as dynamic fission because of its vanishing fission barrier and short time scale, has recently been investigated in various systems \cite{Bocage2000,Filippo2005,Filippo2012,Pagano2018,Piantelli2020}. In various reactions from 25 to 50 MeV/u, the angular distribution of the fission axis w.r.t. the separation direction of projectile-like fragments (PLF) against the target-like fragments (TLF) exhibits an alignment enhancement at very forward angles \cite{Bocage2000}. Similarly, forward-backward asymmetry in the angular distribution has been observed for the dynamic fission events in ${\rm ^{80}Kr + ^{48}Ca}$ at 35 MeV/u using part of FAZIA setup \cite{Piantelli2020,FAZIA2014} and in  ${\rm ^{124}Sn + ^{64}Ni}$ at 35 MeV/u using CHIMERA setup \cite{Filippo2005,Pagano2001}. Besides, the isospin content of the emitted IMFs in the dynamic breakup of PLF shows a slight evolution as a function of the rotating angle of the fissioning PLF \cite{Piantelli2020}.  The short time scale of the dynamic fission is also demonstrated by analysis of the correlation functions of  IMF-IMF pairs \cite{Pagano2018}.

The topic of fast fission with the simultaneous emission of the particles deserves further investigations because the fissioning system provides an appropriate laboratory to probe the density behavior of nuclear symmetry energy, which is of great interest in both nuclear physics and astrophysics  \cite{Liba2021}, particularly after the discovery of the neutron star merging event GW170817.  According to transport model simulations on ${\rm Ar+Au}$ reactions,  the occurrence of fast fission of the heavy target-like fragment (TLF) enhances the emission of the coalescence-invariant neutrons (CIN) in comparison to that in the events without fission\cite{Wuqh2020}. Such enhancement can be associated with the presence of neck emission as shown experimentally \cite{Filippo2012, WRS2014}.  Due to the enriched emission of neutrons, either free or bounded in clusters, and to the accumulation of the isospin effect in the long-time process, the relative neutron richness of the emitted particles provides an effective probe to the symmetry energy. The experimental attempt to extract the nuclear symmetry energy parameter has been reported in inverse kinetic reactions ${\rm ^{124}Sn+^{64}Ni}$, where the dynamic fission of the heavy PLF has been experimentally investigated by measuring the three bodies coincidence.  It has been found that the intermediate-mass fragments emitted from the neck region are more neutron-rich. Comparison to stochastic mean-field (SMF) simulations favors symmetry energy with the slope parameter $L\approx 80$ MeV \cite{Filippo2012}.  Later in ${\rm Ar+Au}$ reactions at 30 MeV/u, the angular evolution of the neutron excess of the emitted light charged particles (LCPs) has been used to probe the symmetry energy, yielding a softer symmetry energy $L\approx 47\pm14$ MeV at 95\% confidential level \cite{zy2017}. 

In this paper, we present the experimental results of the fast fission following Ar+Au reactions at 30 MeV/u using the Compact Spectrometer for Heavy Ion Experiment (CSHINE) \cite{fenhai2021,yijie2021}. The kinetic properties of the fission fragments have been investigated, and the coincident light charged particles (LCPs) at midrapidity have been measured. The angular evolution of the neutron richness of the LCPs is discussed in different linear momentum transfer windows. The article is arranged as follows: Section II presents the experimental setup and the detector performance, Section III introduces the data analysis method on the fission events. The main results are presented in Section IV, and Section V is the summary.

\section{Experimental Setup and Detector Performance}
The experiment was carried out on the Compact Spectrometer for Heavy IoN Experiments (CSHINE) installed in the large chamber at the final focal plane of the Radioactive Ion Beam Line I in Lanzhou (RIBLL1), China. The gold target with a thickness of 1 mg/cm$^2$ was bombarded by the 30 MeV/u $^{40}$Ar beam delivered by the Heavy Ion Research Facility (HIRFL). The beam intensity was about 6 nA.

\begin{figure}[htbp]
	\centering
	\includegraphics[width=0.5 \textwidth]{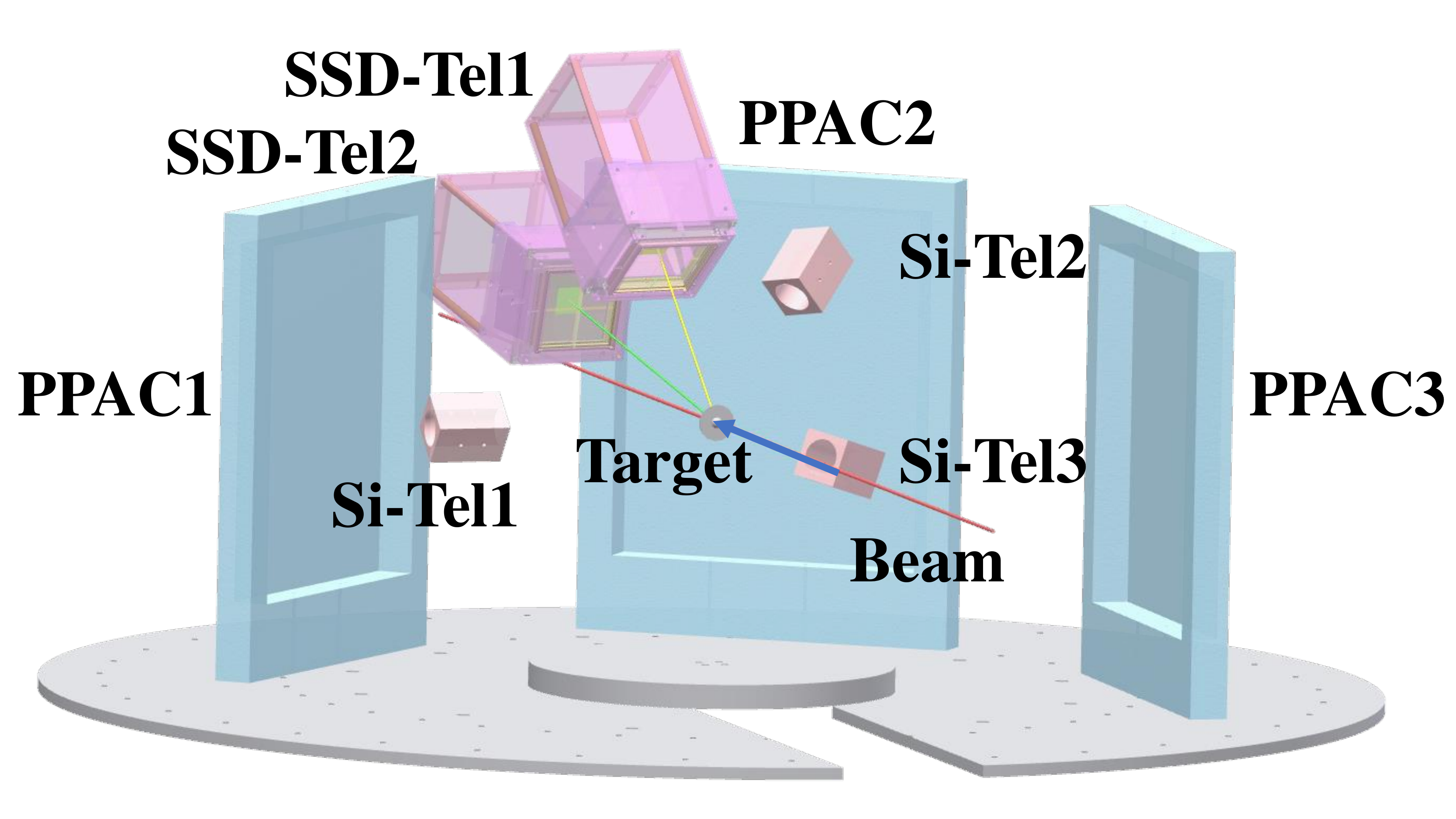}
	\caption{(Color online) Schematic view of the CSHINE detection system in Phase-I.}
	\label{cshine}
\end{figure}

\begin{figure}[htbp]
	\centering
	\includegraphics[width=0.5 \textwidth]{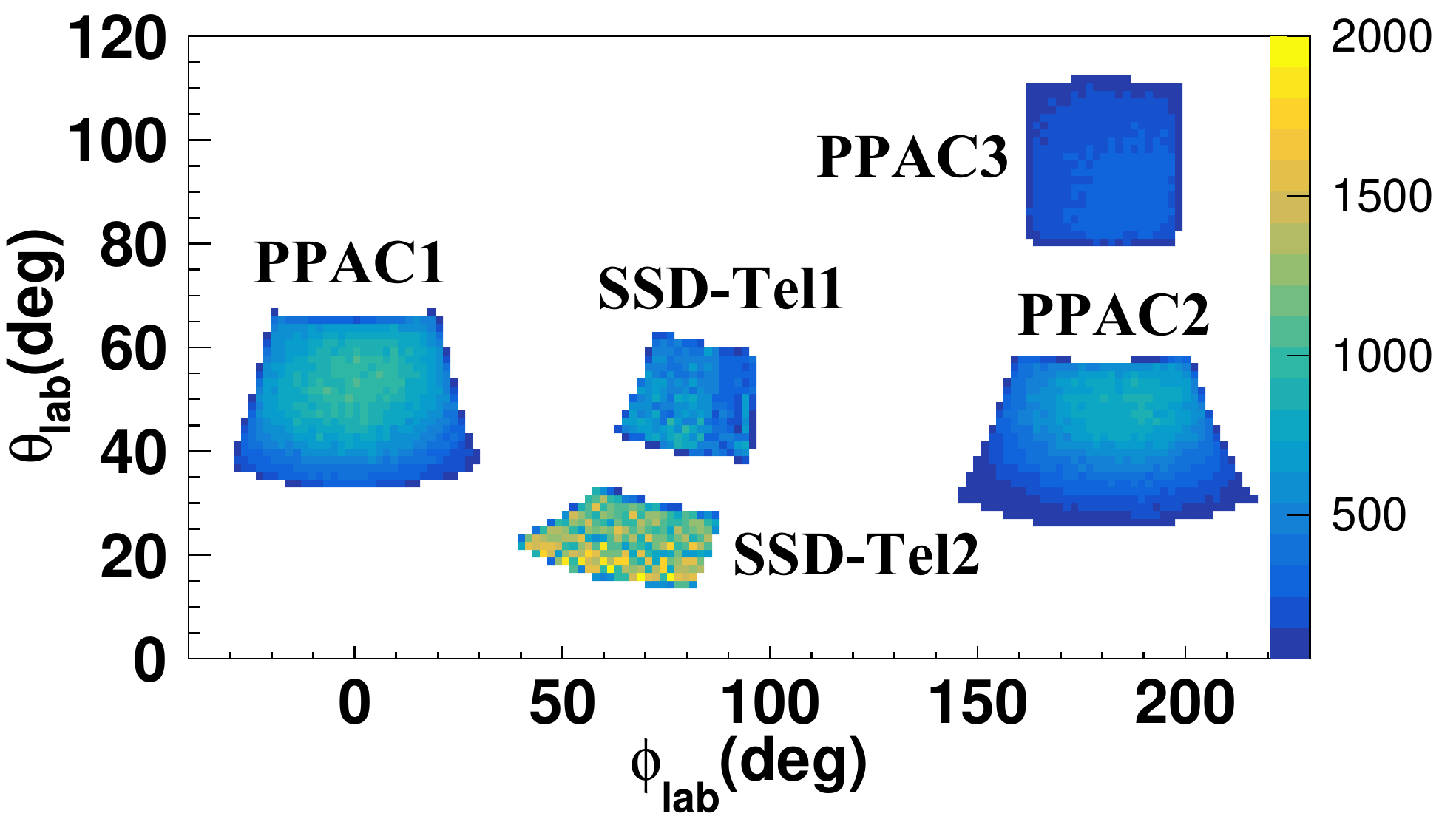}
	\caption{(Color online) The angular coverage of the CSHINE detectors in the laboratory reference. The abscissa is $\phi$ while the ordinate is $\theta$ in the laboratory frame. The PPACs and SSD telescopes are all marked individually.}
	\label{coverage}
\end{figure}

The CSHINE with complete configuration consists of six silicon strip detector (SSD) telescopes and three parallel plate avalanche counters (PPACs) for measuring the light charged particles and the fission fragments (FFs), respectively \cite{fenhai2021}. In phase I, two SSD telescopes and three PPACs were mounted for the beam experiment, as shown in Fig. \ref{cshine}. The PPACs with an active area of $240 \times 280$ mm$^2$ were installed at the distance of 427.5 mm to the target, delivering the position and timing information of the FFs. The PPACs were operated with 4.5 mbar isobutane at 465 V so that the LCPs and intermediate-mass fragments (IMFs) were suppressed. The main PPAC is concentrated at $\theta_{\rm lab}= 50^{\circ}$ and the other two PPACs are placed on the other side of the beam at $\theta_{\rm lab}= 40^{\circ}$ and 95$^{\circ}$, respectively. Each SSD telescope consists of two layers of SSDs for energy loss ($\Delta E$) measurement and one layer of CsI hodoscope with $3\times 3$ units read out by photodiodes (PD) for residual energy $E_{\rm CsI}$ measurement. The first layer of SSD is a 32-strip one-dimensional detector with a thickness of 65 $\mu$m. The second layer,  1500 $\mu$m in thickness, is a two-dimensional detector with 32 strips on both sides. The strips are all 1.9 mm wide, and the inter-strip distance is 0.1 mm.  The geometry coverage of the PPACs and the SSD telescopes in the laboratory reference is presented in Fig. \ref{coverage}. It is shown that the angular range of $\theta_{\rm lab}= 18^{\circ}-60^{\circ}$ is partially covered for the measurement of the charged particles. In addition, three small silicon telescopes consisting of two Si (Au) barrier detectors and a CsI unit are placed at large angles to measure the evaporation LCPs. The geometric parameters, including the distance from the center of each detector to the target $d$, the polar angle $\theta$, the azimuth angle $\phi$, and the sensitive area $S$ of the detectors of CSHINE components are listed in Table 1.

\begin{table}[htbp]
	\label{tab:cee_para}
	\caption{Geometric parameters of PPACs, SSD telescopes and small Si(Au) telescopes.}
	\begin{center}
		\begin{tabular}
			{ccccc}
			\hline 
			Detector & $d(mm)$ & $\theta^{\circ}$ & $\phi^{\circ}$ & $S(mm^2)$ \\
			\hline
			SSD-Tel1  & 161.9 & 50.7 & 81.7 & 64$\times$64\\
			SSD-Tel2  & 221.9 & 22.3 & 64.5 & 64$\times$64\\
			PPAC1  & 427.5 & 50 & 0 & 240$\times$280\\
			PPAC2  & 427.5 & 40 & 180 & 240$\times$280\\
			PPAC3  & 427.5 & 95 & 180 & 240$\times$280\\
			Si-Tel1  & 185.5 & 90 & 0 & $16\pi$ \\
			Si-Tel2  & 145.5 & 130 & 40 & $16\pi$\\
			Si-Tel3  & 165.5 & 130 & 15 & $16\pi$\\
			\hline
		\end{tabular}
	\end{center}
\end{table}
\vspace*{-2mm}

\begin{figure}[htbp]
	\centering
	\includegraphics[width=0.5 \textwidth]{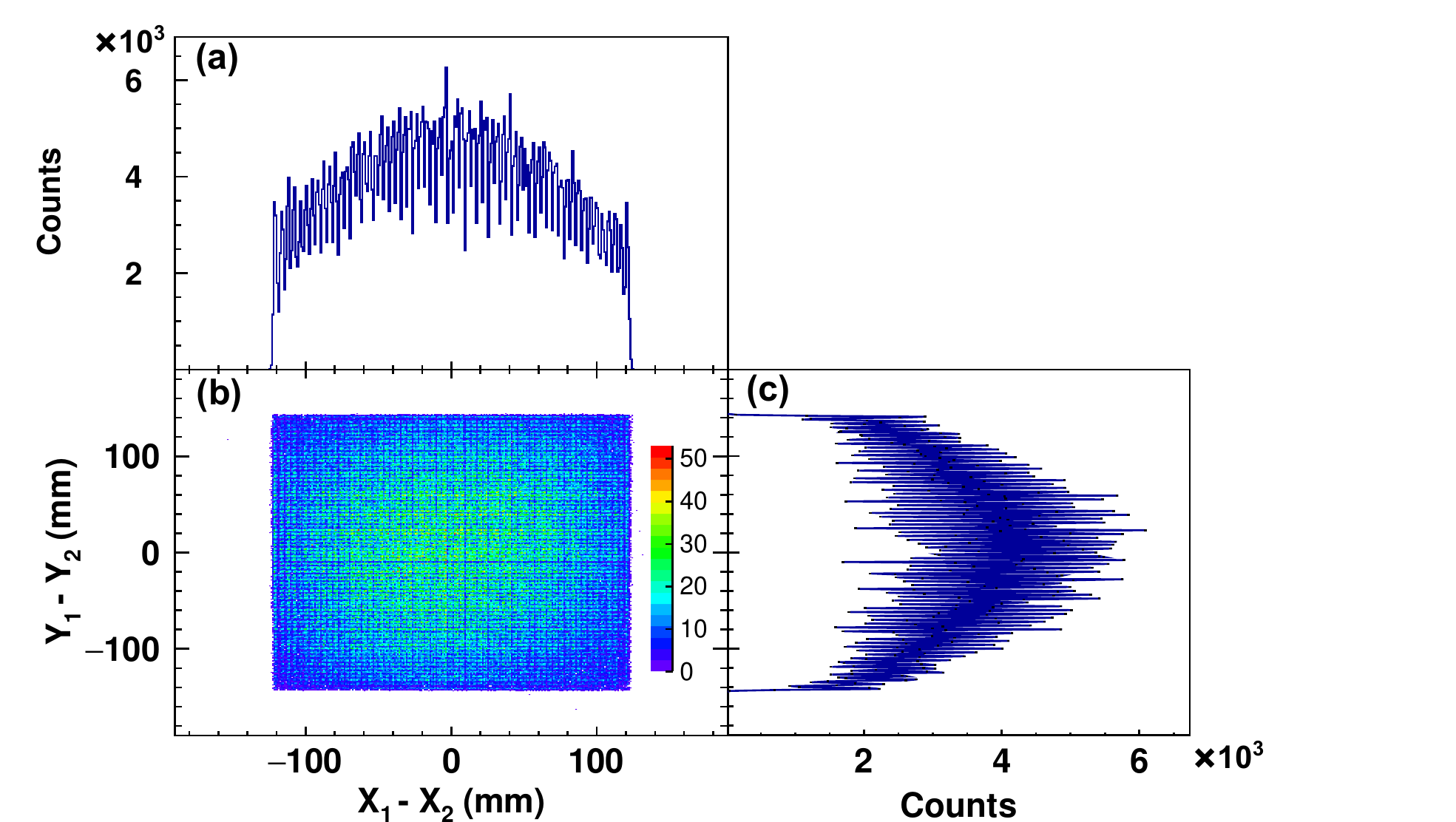}
	\caption{(Color online) The performance plots of PPAC1. Panel (b) is a two-dimensional distribution of $Y_1-Y_2$  versus $X_1-X_2$ . The projection to X1-X2 and Y1-Y2 dimensions are shown in panel (a) and (c), respectively.}
	\label{ppacs}
\end{figure}

PPAC is a multi-wire chamber working in a region of limited proportionality. The signals induced by the incident fragments on an individual wire of the anode plane, either X or Y, are transferred through a delay line to both ends. The time delay of the two signals $X_1$ and $X_2$ ($Y_1$ and $Y_2$)  with respect to the signal collected on the cathode plane, which delivers the timing information,  gives the X (Y) position of the hit in the sensitive area. Fig. \ref{ppacs} (b) shows a two-dimensional histogram of $Y_1-Y_2$ $vs$ $X_1-X_2$  for PPAC1 as an example. The projections to X and Y direction are plotted in Fig. \ref{ppacs} (a) and (c), respectively. A good performance in timing, corresponding to good position resolution, manifests itself in the sharp boundary for the two-dimensional distribution and the well-separated individual peaks on the projections. The distance of the neighboring wires is 4 mm, and there are 61 peaks and 71 peaks in Fig. \ref{ppacs} (a) and (c), respectively. The time resolution of $\sigma_{\rm T} = 300$ ps and the position resolution of $\sigma_{\rm r}=1.35$ mm can be derived from the data. The mechanical structure and the overall performance of the PPACs can be found in  \cite{wei2020}.

\begin{figure}[htbp]
	\centering
	\includegraphics[width=0.5 \textwidth]{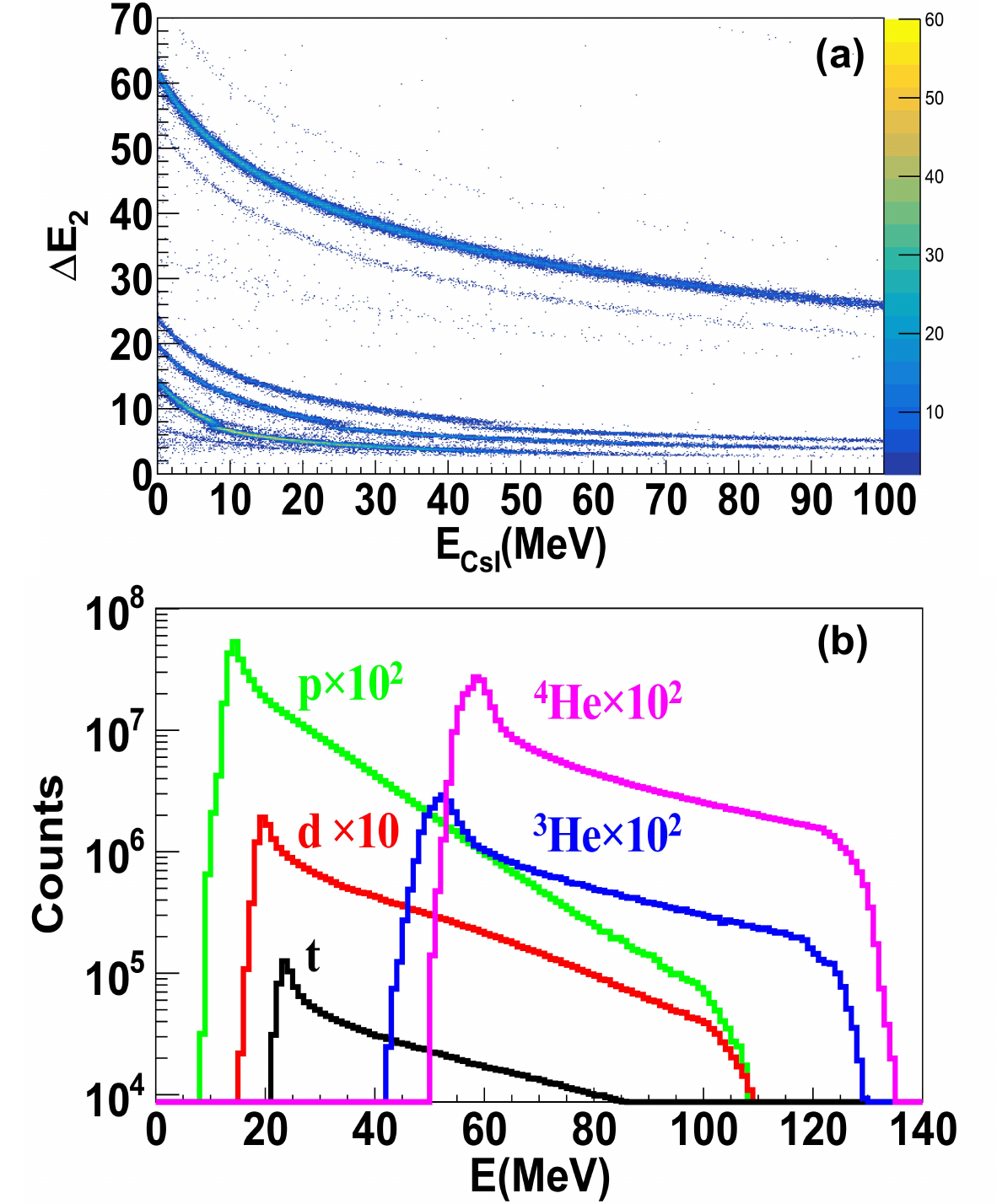}
	\caption{(Color online) SSD telescopes performances in the experiment. (a) $\Delta{E_2}$-E$_{CsI}$ scattering plot of SSD telescope 2. (b) The energy spectra of p, d, t, $^3$He and $^4$He, respectively.}
	\label{ssds}
\end{figure}

The SSD telescopes are advantageous for their high granularity and high energy resolution. Owing the 2 mm width strip, one obtains the angular resolution better than $1^\circ$ at the current geometrical positions. The $\Delta E_1$ and $\Delta E_2$ silicon strip detectors are calibrated using the precise pulse generator and the $^{239}$Pu $\alpha$ source. After the SSD calibration is done, the CsI hodoscope can be calibrated for each individual isotope from  $\Delta E_2-E_{\rm CsI}$ scattering plot, where the energy deposit in CsI can be calculated using LISE++ package \cite{lise2016}. The total energy resolution of the SSD telescopes is about 2$\%$ evaluated by the Monte-Carlo simulation studies reproducing the trend and the width of all isotope bands on the $\Delta{E_2}-E_{\rm CsI}$ plot \cite{fenhai2021}. The performance of the SSD telescopes in CSHINE is summarized in Fig. \ref{ssds}. Panel (a) presents the $\Delta{E_2}$-E$_{\rm CsI}$ scattering plot of one CsI unit in the SSD telescope 2 as an example. It is shown that the isotopes of hydrogen and helium are clearly identified. From the width of each band, the mass resolution of the LCPs, $\Delta M/M=1$,  is estimated.  Since the high voltage for the first layer of the SSD telescopes failed for most of the time in the experiment and caused poor particle identification on $\Delta{E_1}-\Delta{E_2}$ plots, only the high-energy LCPs penetrating through $\Delta{E_2}$ SSD were analyzed in current studies.  Fig. \ref{ssds} (b) presents the energy spectra of p, d, t, $^3$He, and $^4$He in coincidence with the FFs, respectively. See Ref. \cite{fenhai2021} for the structural framework, performance, and detailed analysis of the SSD telescopes.

In addition to the inclusive trigger for all detectors, three types of coincident trigger signals are constructed in this experiment. i) The two-fission-fragment event in the PPACs, ii) The two-body LCP event in the SSD telescopes, and iii) One LCP coincided with two FFs in PPACs. For the measurement of two FFs, while PPAC1 is served as the primary FF detector, PPAC2 or PPAC3 on the other side of the beamline provides the coincident FF. The circuit diagram of the trigger electronics in the experiment can be found in \cite{yijie2021}.

In the 120-hour beam experiment, the inclusive event of at least one LCP is about $1.6 \times 10^7$  after applying a fraction divide factor 50, the event of two fission fragments is about $5 \times 10^6$, and for the event of two fission fragments plus one light charged particle, the statistics is about $3.64\times 10^5$.

\section{Data Analysis Method}

\begin{figure}[htbp]
	\centering
	\includegraphics[width=0.48\textwidth]{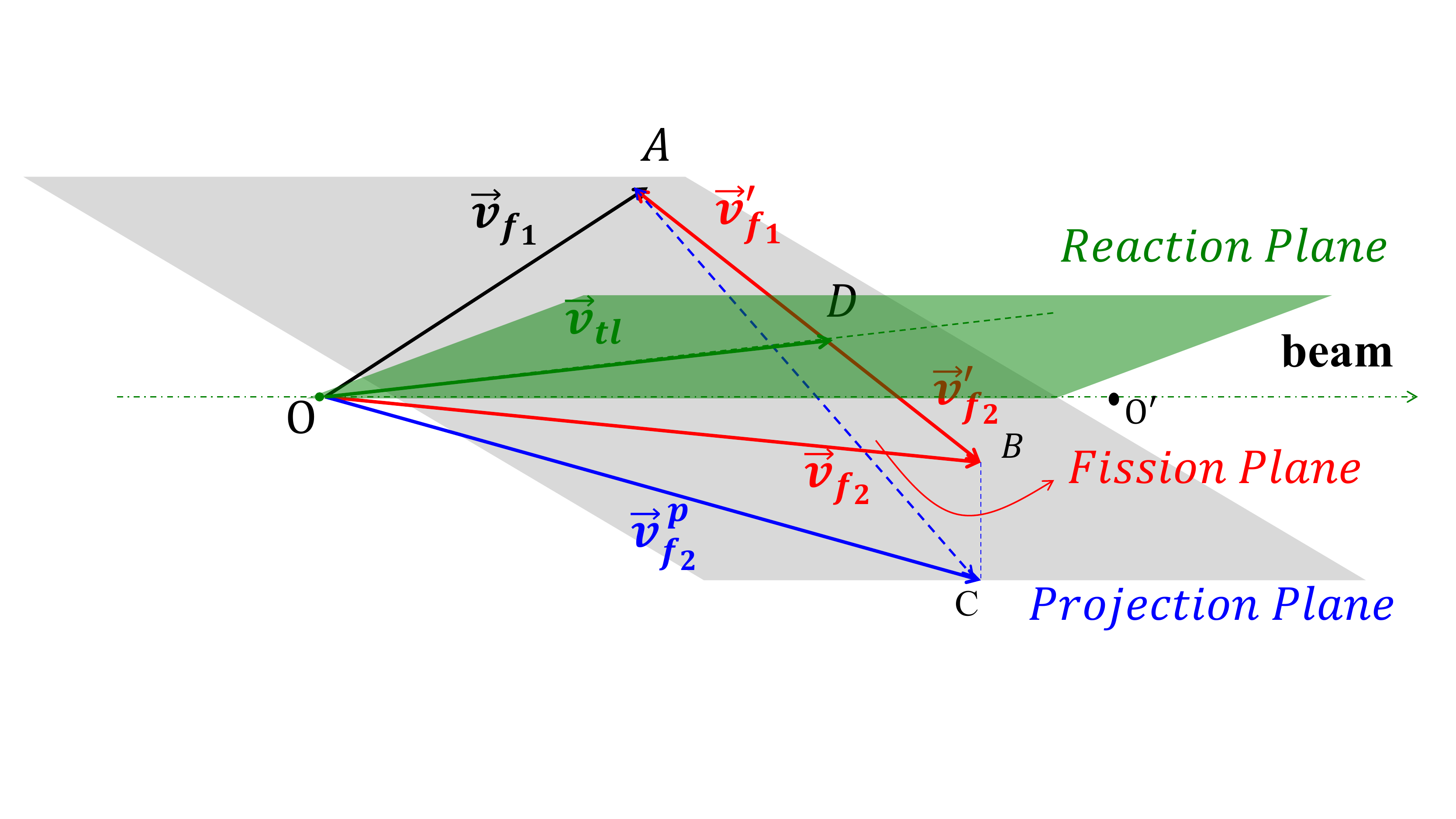}
	\caption{The geometric diagram of the velocities of the two TLF-fission fragments.}
	\label{fission_vector}
\end{figure}

In the picture of incomplete fusion, a heavy target-like fragment (TLF) is formed by the fusion of part of the projectile and the target nuclei. The fraction of the total momentum of the projectile transferred to the TLF is called linear momentum transfer (LMT). With a certain probability depending on the system's total angular momentum, the TLF may undergo fission or fast fission in competition with the emission residue channel. For the fission events, Fig. \ref{fission_vector} presents the kinetic geometry of the TLF fission event. As shown, the origin point O is the target nuclei in the laboratory system, the vector OO$^{'}$ represents the direction of the beam. The velocity vectors $\vec{v}_{\rm f_1}$ and $\vec{v}_{\rm f_2}$ of the two fission fragments in the laboratory system are represented by OA and OB.   $\vec{v}_{\rm tl}$ is the velocity of the TLF, and the velocities of the two fragments in the center-of-mass system of the fissioning TLF are represented by $\vec{v}^{\,'}_{\rm f_1}$ and $\vec{v}^{\,'}_{\rm f_2}$ sitting back-to-back collinearly. Here, we define plane OAB as the fission plane, plane OO$^{'}$A as the projection plane, and the reaction plane is defined as plane DOO$^{'}$.

In the experiment,  the position and energy information of the LCPs can be measured accurately. For the fission fragments, however, because the start timing of the reaction can not be measured with sufficient precision, the absolute time-of-flight (TOF) is not available. Only the difference of TOF between the two fission fragments can be derived with sub-nanosecond timing resolution for each detector, which reaches 300 ps in the source test \cite{wei2020}. Besides, the direction of the velocity vectors can be precisely determined by the hit positions recorded by the PPACs. 

\begin{figure}[htbp]
	\centering
	\includegraphics[width=0.45\textwidth]{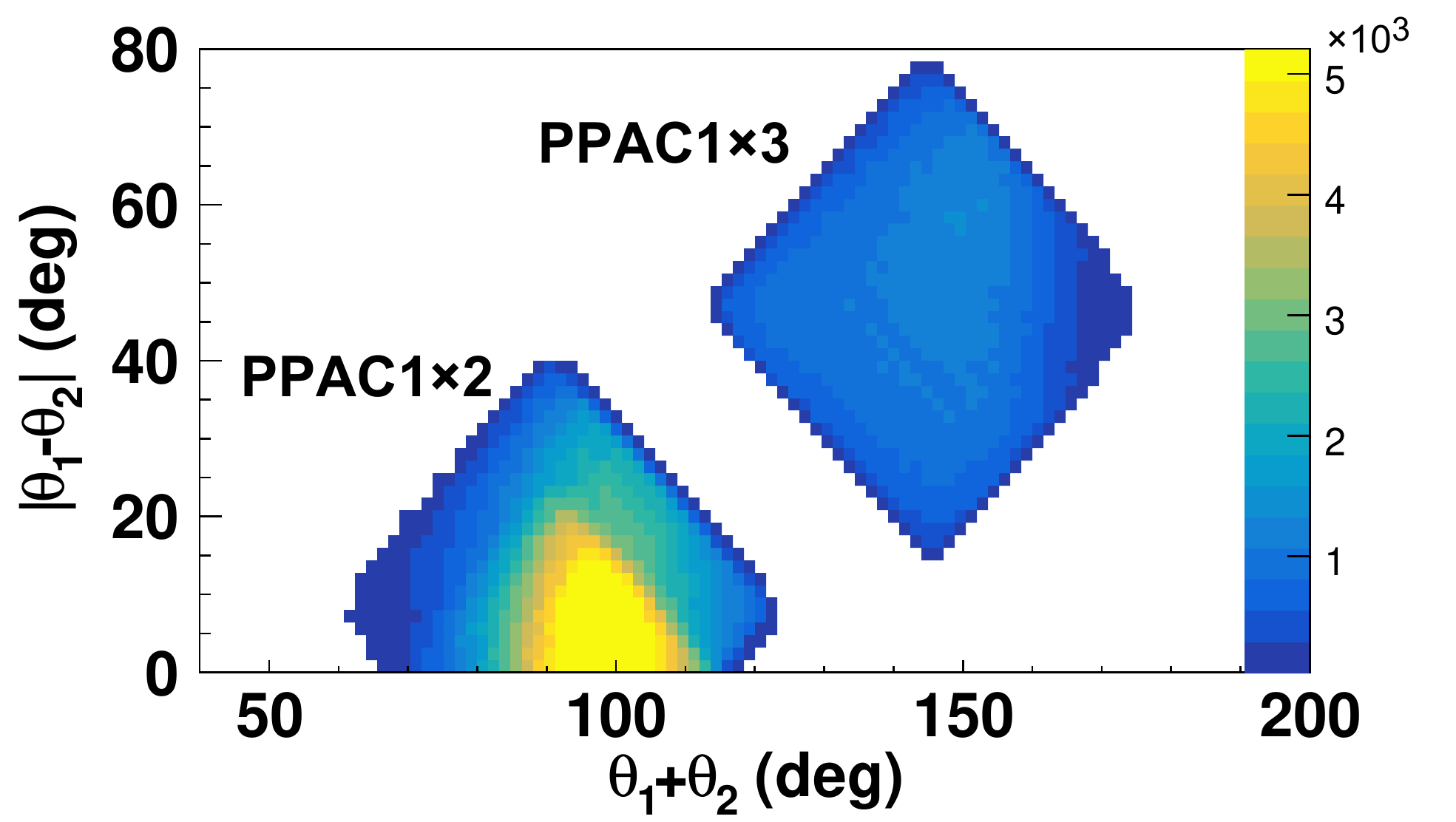}	
	\caption{(Color online) The experimental correlation between $|\theta_1-\theta_2|$  vs. $\theta_1+\theta_2$ of  two fission fragments marked  by subscribes 1 and 2, respectively.}
	\label{angle_corr}
\end{figure}
\begin{figure}[htbp]
	\centering
	\includegraphics[width=0.5\textwidth]{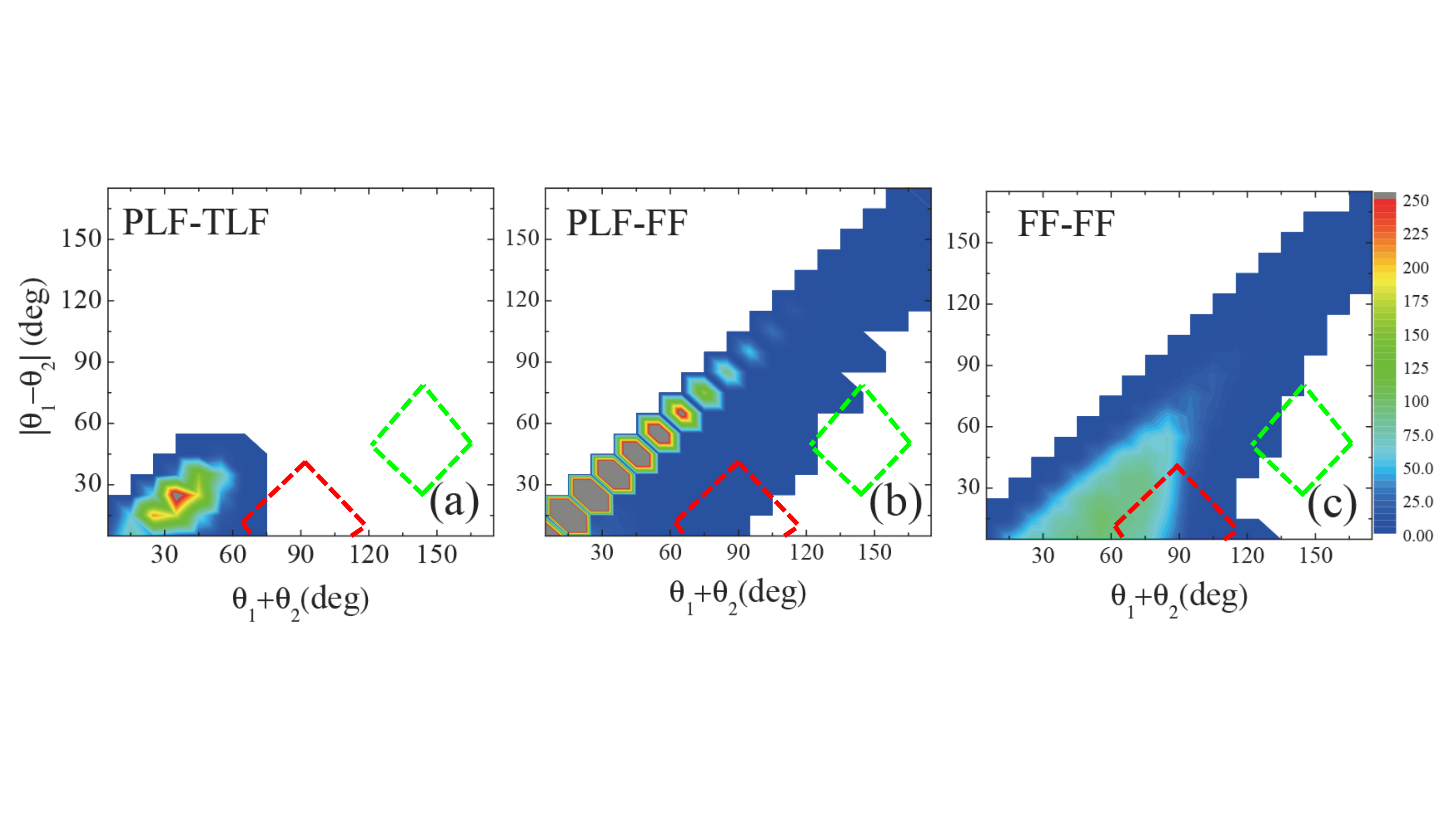}	
	\caption{(Color online) The correlation between $|\theta_1-\theta_2|$  vs. $\theta_1+\theta_2$ of  two fission fragments  in ImQMD calculations.   Panel (a-c) presents the results for PLF-TLF events, PLF-FF events and FF-FF events, respectively. The polygons in the panels indicate the experimental coverage.}
	\label{angle_iqmd}
\end{figure}

It is important to assure that the coincident events recorded by  PPAC1 with PPAC2 (marked by ${\rm PPAC 1\times 2}$) or with PPAC3 (marked by ${\rm PPAC 1\times 3}$) are the fission fragments (marked by FF-FF) since the two-body correlation is possibly contaminated by the correlation of the PLF with TLF (PLF-TLF)  or the correlation of the PLF with one of the fission fragments from the TLF (PLF-FF).  First, the high voltage is set that the response of PPAC to PLF is highly suppressed, given that the energy loss of PLF is lower than FFs by more than one order of magnitude. Second, the geometric coverage of the PPACs is out of the angular range of the PLF.  In order to verify this point, we check the angular correlation of the two fragments in comparison with the prediction of the improved quantum molecular dynamics (ImQMD) transport model calculations. The details of the ImQMD model can be found in \cite{Wuqh2020}. Fig. \ref{angle_corr} presents the experimental correlation between the polar angles of the two fragments on the scattering plot of $|\theta_1-\theta_2|$  vs  $\theta_1+\theta_2$ . It is shown that the events of \pacab~  are well separated from those of \pacac. 
For comparison, the results of transport model simulations are displayed in  Fig. \ref{angle_iqmd} (a-c) in the form of $|\theta_1-\theta_2|$  as a function of $\theta_1+\theta_2$. The calculations from central to peripheral reactions ($b=4-10$ fm) are accumulated according to the corresponding weights. The experimental coverage of \pacab~ and \pacac~ are indicated by the red and green dashed polygons. Panels (a), (b), and (c) represent the  PLF-TLF,  PLF-FF, and FF-FF correlations in ImQMD simulations, respectively.  It is shown that the angular correlation of the PLF-TLF events is mostly out of the experimental coverage. For the reactions undergoing fission, the PLF-FF correlations are partially covered by the PPAC1 and PPAC2 with a tiny portion, as indicated by the marked area's colorful contour compared with the most populated area in panel (b). It indicates that the coincident events of two heavy fragments are dominated by the FF-FF events, as confirmed in panel (c).


\begin{figure}[htbp]
	\centering
	\includegraphics[width=0.48\textwidth]{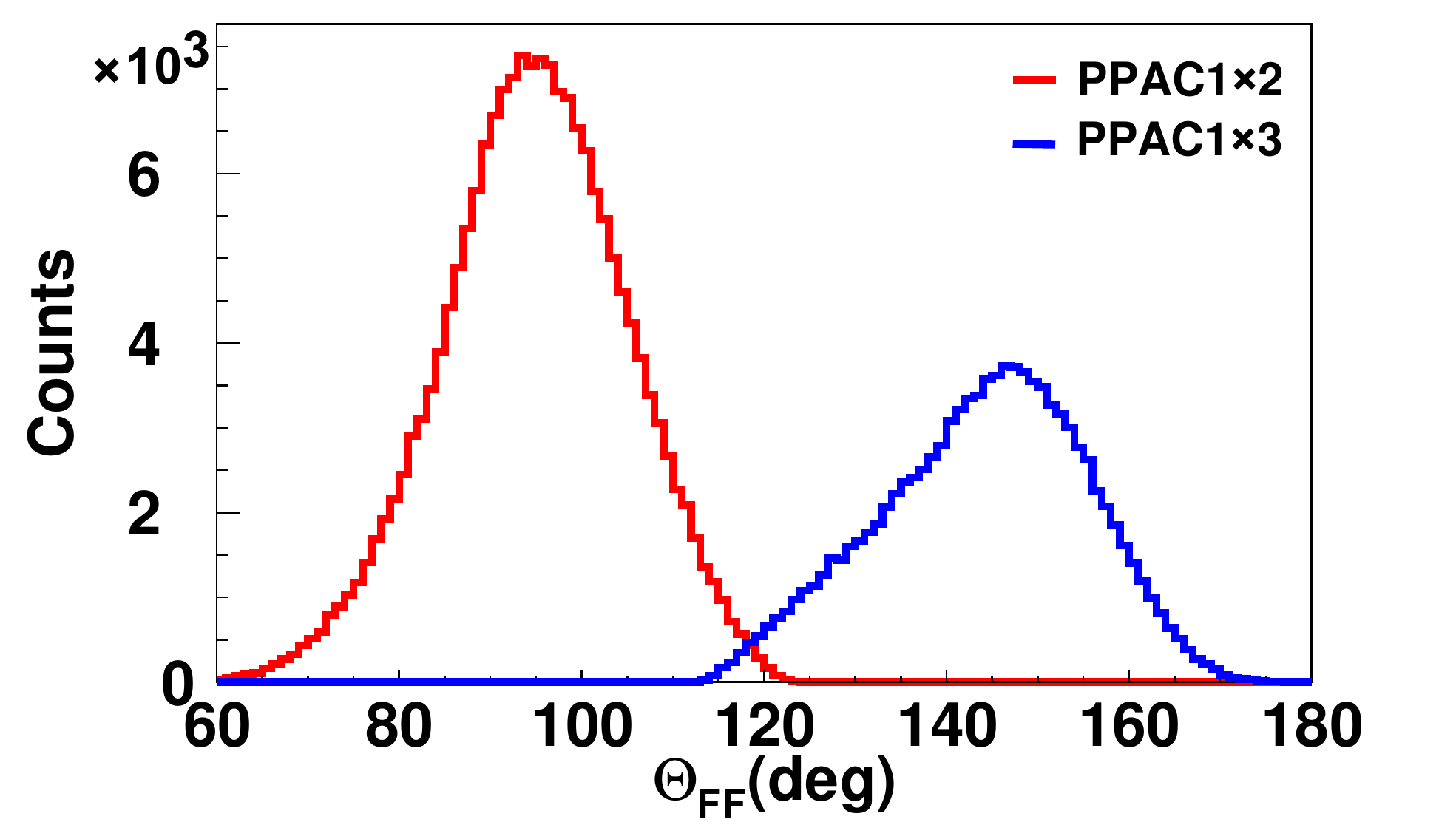}
	\caption{The folding angle distribution of two-body fission events. The red histogram represents the events of ${\rm PPAC 1\times 2}$ and the blue histogram represents the ${\rm PPAC 1\times 3}$ events.}	
	\label{theta_12}
\end{figure}

Once the events with two FFs are identified, the folding angle method has been used to measure the LMT of the reaction. As shown in Fig. \ref{fission_vector}, the folding angle $\Theta_{\rm FF}$ is defined as the angle $\angle AOC$,  spanned by the projection of the velocity vectors of the  FFs on the projection plane. Clearly, given the relative velocity $\Delta v_{\rm 12}$ of the FFs, the folding angle depends on the velocity of the TLF $v_{\rm tl}$, i.e., the larger $v_{\rm tl}$, the smaller  $\Theta_{\rm FF}$. Fig. \ref{theta_12} presents the distribution of the folding angle $\Theta_{\rm FF}$. The coincident events of \pacab~ distribute in the range between $60^{\circ}-120^{\circ}$ with the peak situating at about $90^{\circ}$, corresponding to a larger LMT (red), while the  \pacac~ events sit in the range of $110^{\circ}-170^{\circ}$  corresponds to a smaller LMT (blue).  The valley between the two histograms is due to the deficiency caused by the gap between PPAC2 and PPAC3. 

\begin{figure}[htbp]
	\centering
	\includegraphics[width=0.48\textwidth]{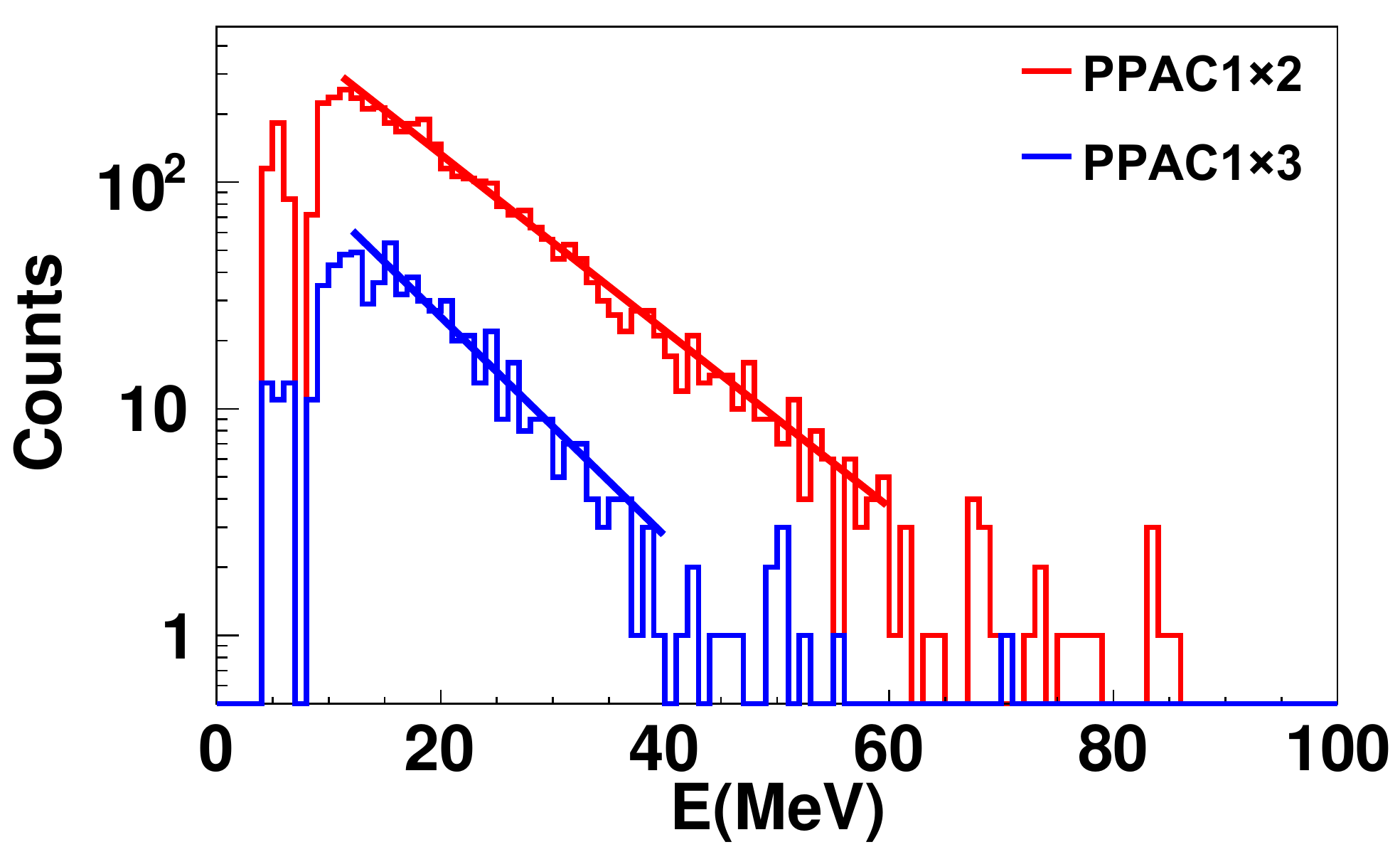}
	\caption{The energy spectra of protons in the Small Si-Tel3 in coincidence with the  ${\rm PPAC 1\times 2}$ (red) and ${\rm PPAC 1\times 3}$ (blue) events.  The straight lines are the logarithmic fitting to the descents of the energy spectra.}
	\label{energy_spectra}
\end{figure}

One can further check the energy spectra of LCP from the statistic emission in the two groups of events to make sure the experimental folding angle is a measure of the LMT. Model independently, a larger LMT causes higher excitation energy of the TLF, for which a larger slope temperature parameter of the emission LCP can be expected. To check this scenario, one can simply survey the energy spectra of evaporation particles at very large angles, where only the evaporation source dominates. Figure \ref{energy_spectra} presents the energy spectra of protons recorded in the small Si3 installed at $\theta_{\rm lab}=130^\circ$ in coincidence with ${\rm PPAC 1\times 2}$ (red)  and ${\rm PPAC 1\times 3}$ (blue) fission events. Because of the low statistics, finer division of the $\Theta_{\rm FF}$ is not achievable. The descents of the energy spectra are fitted by a logarithmic function $Y=a\exp{\left(-E/T_{\rm sl}\right)}$ without taking into account the kinetic energy difference and the Coulomb barrier difference of these two groups of event, which mainly influence the peak of the spectra.  The slope parameter $T_{\rm sl}= 6.2\pm 0.2$ and $5.3\pm 0.2$ MeV are deduced from the fitting for the two groups of events, respectively. Evidently, the slope parameter is larger in the group of ${\rm PPAC 1\times 2}$ events than that in  ${\rm PPAC 1\times 3}$. It is consistent with the picture that the folding angle of the FFs reflects the LMT of the reaction. Namely, with a smaller folding angle, or larger LMT, the TLF formed in the incomplete fusion gains higher excitation energy.

In order to reconstruct the complete kinetics of the fission of TLF, one has to measure the velocity vectors of the two FFs. It requires to measure both the time and the distance of flight of the FFs. However, the start timing signal, which is supposed to be delivered by the cyclotron's radio frequency (RF) signal, was not available in the experiment. Thus, to obtain the velocity, one relies on the Viola's systematics describing the velocity of the FFs in the center of mass reference of the fissioning nucleus. For symmetric fission, the kinetic energy of fission fragments is mainly supplied by the Coulomb energy at the scission point, and the most probable velocity of each FF is 1.2 cm/ns \cite{viola}. For the fast fission that occurred in heavy-ion reactions at Fermi energies, recent simulations with the ImQMD model verify that the distribution of the relative velocity of the FFs is indeed peaked in the vicinity of 2.4 cm/ns with a broadening less than 10\%. It shows insignificant dependence on the mass asymmetry of the FFs  \cite{Wuqh2019}. Here we neglect the influence of highly-aligned events with the fission axis sitting in parallel with the flying direction of the TLF. These highly-aligned events, in which the relative velocity may violate the assumption as observed at higher beam energy\cite{Bocage2000}, are anyway out of the geometrical reach in the current experimental setup.   Following the Viola systematics, we first assume that relative velocity between the two FFs is fixed at 2.4 cm/ns. Under this assumption, applying cosine theorem to the velocity vector triangle $\Delta OAB$ as shown in Fig. \ref{fission_vector}, one writes

\begin{equation}
	v_{\rm f_1}^2+v_{\rm f_2}^2-2 v_{\rm f_1} v_{\rm f_2}\cos\theta=(v_{\rm f_1}'+v_{\rm f_2}')^2 ~,
\end{equation}
where $\theta$ is the angle met by the two velocity vectors $\vec{v}_{\rm f_1}$ and $\vec{v}_{\rm f_2}$.  Using the TOF difference $\Delta TOF$ and the hit position information on PPACs, one gets

\begin{equation}
	\frac{L_1}{v_{\rm f_1}}-\frac{L_2}{v_{\rm f_2}} = \Delta TOF    ~,
\end{equation}
where $L_{1}$ and $L_{2}$ represent the flight distances of the two FFs deduced by the distance from the hit positions on PPACs to the target for the two FFs, respectively. The Viola systematics reads

\begin{equation}
	|v_{\rm f_1}'+v_{\rm f_2}'| = 2.4~{\rm cm/ns}   ~.
\end{equation}
With equations $(1-3)$, one can readily reconstruct the velocity of the two FFs numerically. If only the fragments following the fission of TLF are recorded, the distribution of both velocities will exhibit a one-component feature.

\begin{figure}[htbp]
	\centering
	\includegraphics[width=0.48\textwidth]{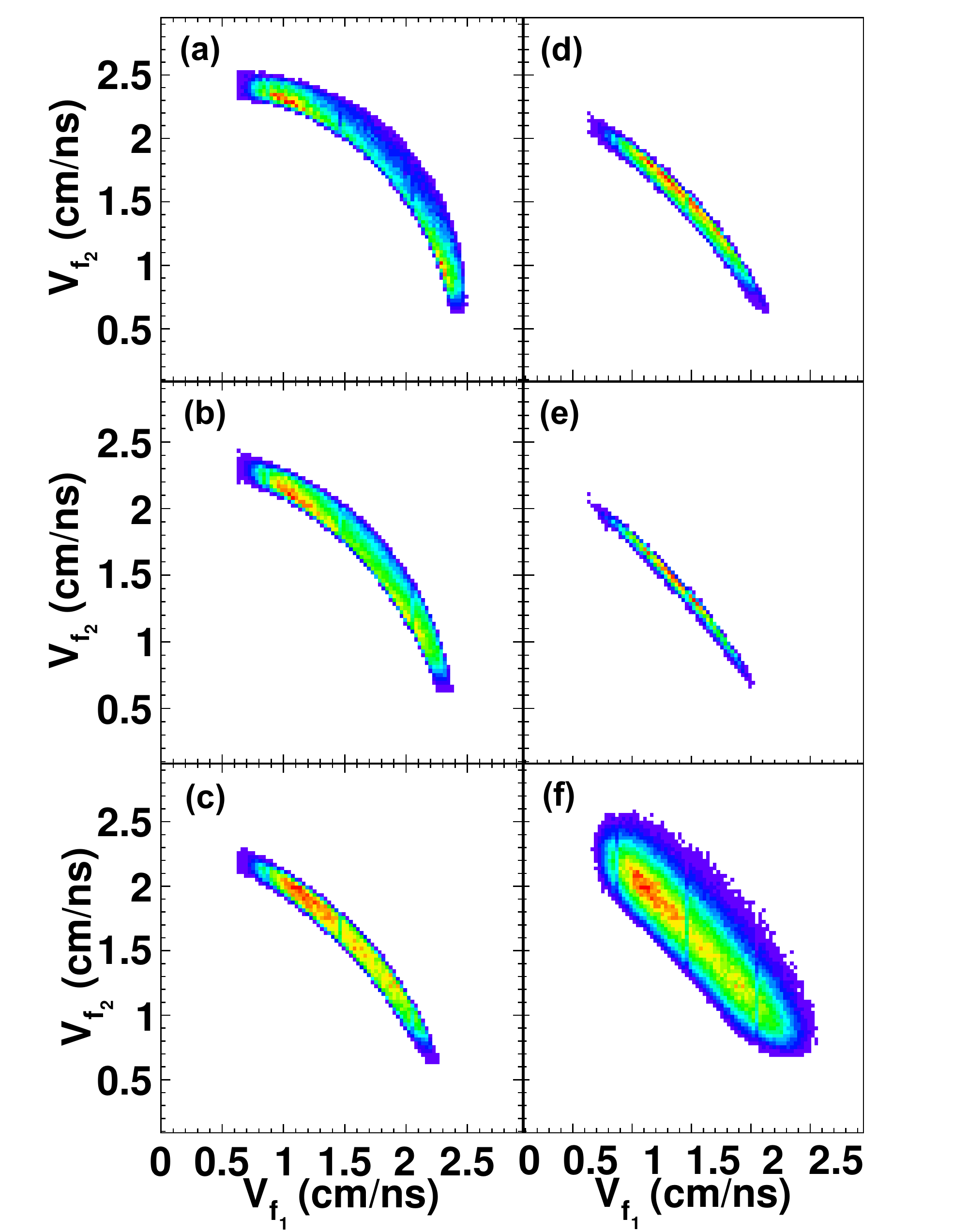}
	\caption{The velocity correlation for ${\rm PPAC 1\times 2}$ fission events. The abscissa is the velocity of the FF firing PPAC1, and the ordinate is the velocity of the FF firing PPAC2.}
	\label{v_corr}
\end{figure}

Fig. \ref{v_corr} presents the correlation of the velocity $v_{\rm f_1}$ and $v_{\rm f_2}$ of the two FFs in laboratory for the  ${\rm PPAC }1\times 2$ fission events. Panels (a-e) represents the correlations in five windows of the folding angle $\Theta_{\rm FF}=75^\circ-85^\circ,~85^\circ-95^\circ,~95^\circ-105^\circ,~105^\circ-115^\circ$ and $115^\circ-125^\circ$, respectively. Panel (f) is the sum of all the events.  Except for the lowest $\Theta_{\rm FF}=75^\circ-85^\circ$ window, where two components are seemingly presented indicating that they are originating from different sources, the anti-correlation is well visible in all other windows, and the kinetic feature of the two FFs exhibits insignificant difference. The velocity correlation for the ${\rm PPAC 1\times 3}$ fission events, which are not presented here, does not show a two-component feature.

\begin{figure}[htbp]
	\centering
	\includegraphics[width=0.5\textwidth]{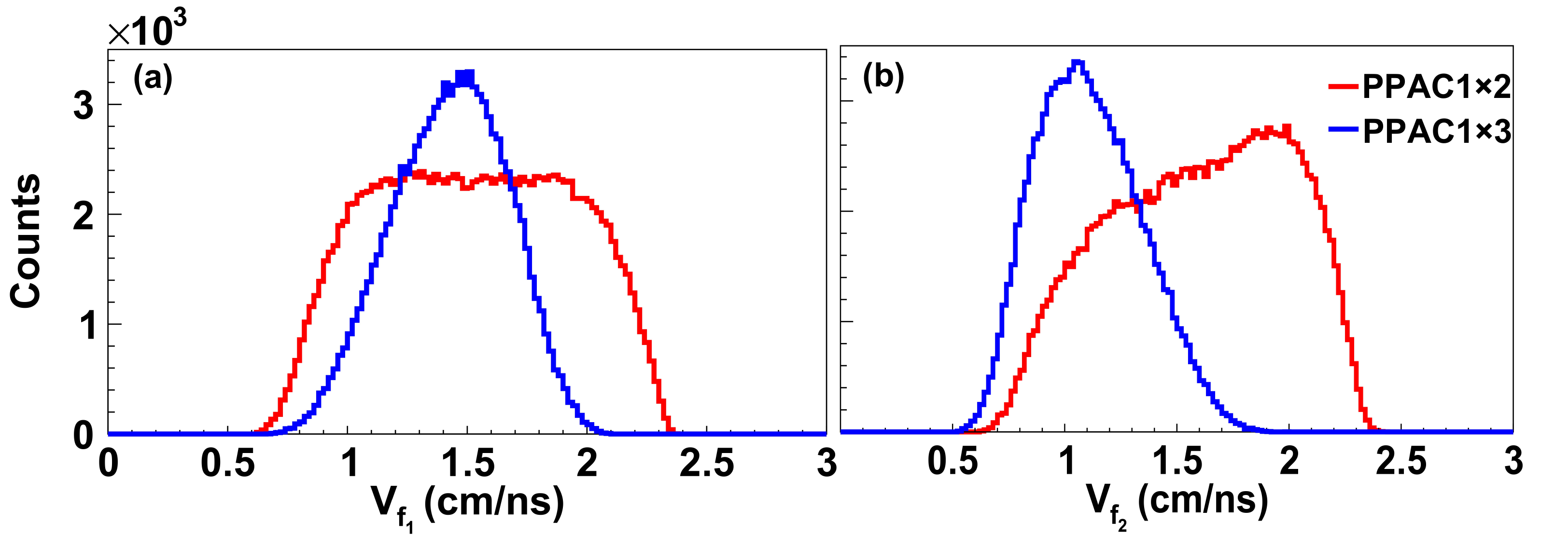}
	\caption{The velocity distribution of the two  FFs in the central (red) and peripheral (blue) reactions. Panel (a) presents $v_{f_1}$ recorded in PPAC1 and panel (b) presents $v_{f_2}$ recorded in PPAC2 (red) or PPAC3 (blue).}
	\label{velocities}
\end{figure}

Fig. \ref{velocities} presents the velocity distributions of the two fission fragments after imposing the union cuts (see Table 2) for both the central collisions (\pacab) and the peripheral collisions (\pacac), respectively. Here, panel (a) presents the velocity $v_{\rm f_1}$ of the FF in PPAC1, while panel (b) is the velocity $v_{\rm f_2}$ of the fragment measured in PPAC2 (red) or PPAC3 (blue), respectively. For  $v_{\rm f_1}$, the broadening of the distribution is $0.5-2.5$ cm/ns in central reactions, larger than the range $0.8-2$ cm/ns as in peripheral reactions. Very similarly, the width of  $v_{\rm f_2}$ is larger in  PPAC2 than that in PPAC3 because the former (later) corresponds to central (peripheral) reactions, respectively. It is as expected that the excitation energy of the TLF is higher in the event group of \pacab~ corresponding to larger LMT, for which a wider distribution of the velocity of the FFs is observed. In peripheral reactions,in addition, a part of the LMT is converted into exciting the collective rotation of the TLF. Detailed analysis on the variance of the velocity will be discussed in the next section.

\begin{figure}[htbp]
	\centering
	\includegraphics[width=0.48\textwidth]{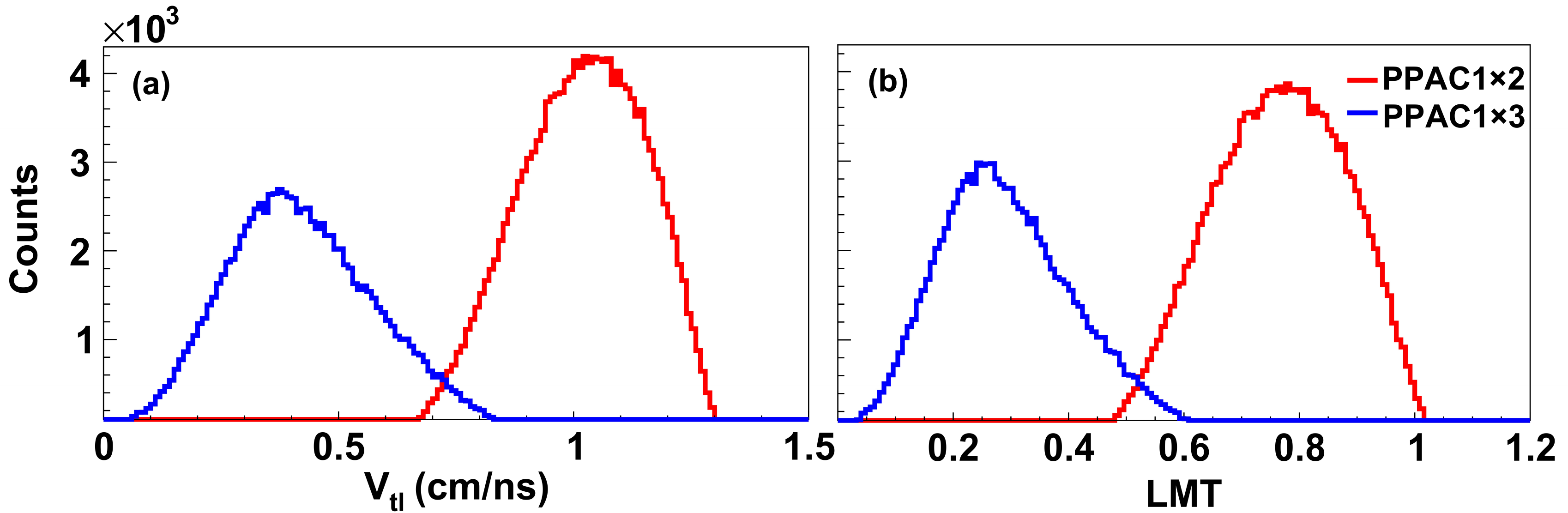}
	\caption{(Color online) The velocity of the TLF (a) and the LMT calculated under the assumption of symmetric fission for \pacab (red) and \pacac fission events  (blue).}
	\label{lmt_vtlf}
\end{figure}

\begin{figure}[htbp]
	\centering
	\includegraphics[width=0.48\textwidth]{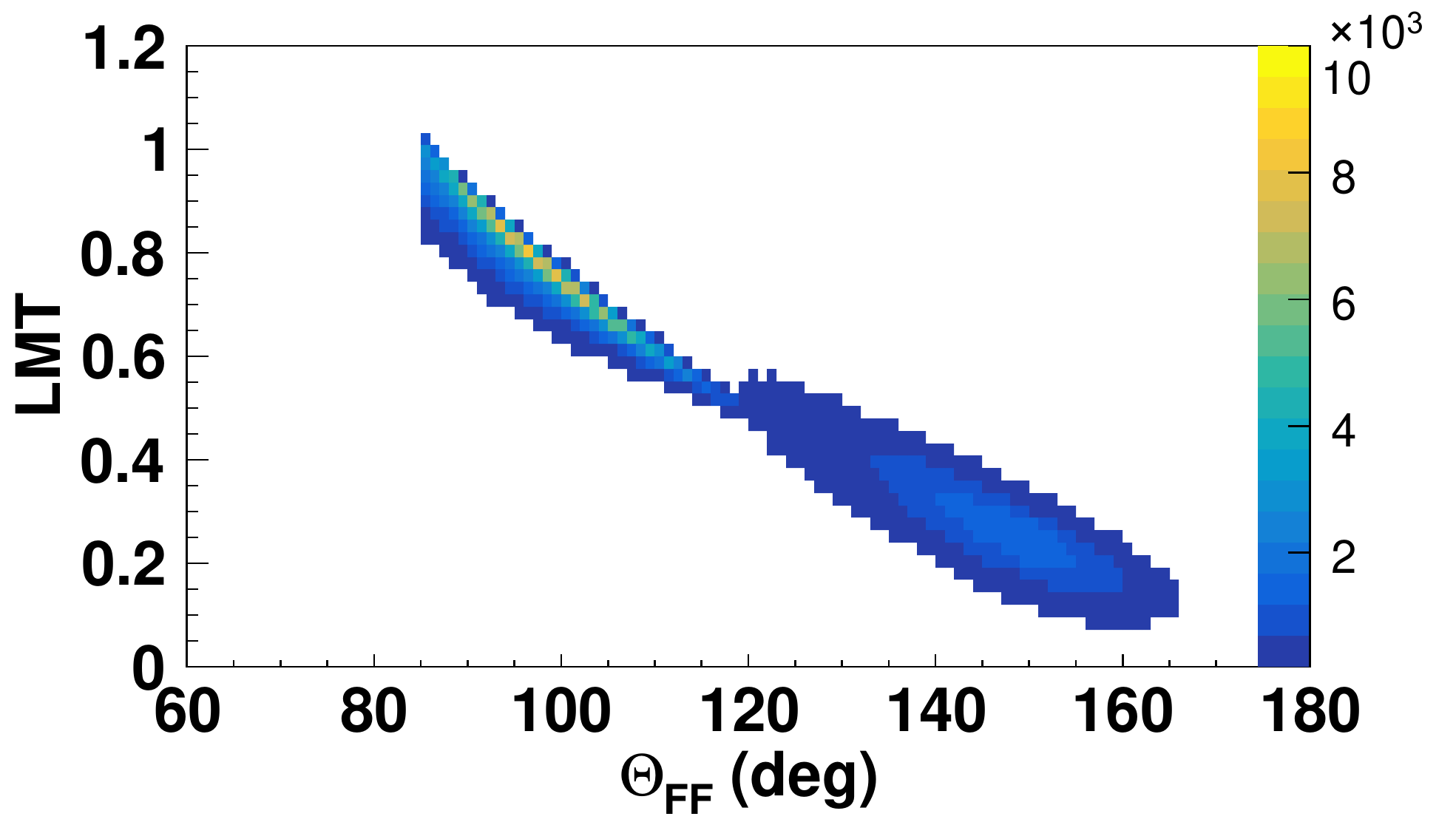}
	\caption{ (Color online)  The correlation between LMT and folding angle $\Theta_{\rm FF}$. }
	\label{lmt_fold_corr}
\end{figure}

As long as the velocities of the FFs are reconstructed using the Viola systematics, the velocity of the TLF in the center of mass frame can be calculated if the mass ratio of the two FFs is known. However, the momentum conservation criteria is already used in equation (2), in order to calculate the $v_{\rm tl}$ and further LMT, one can fix the direction of $\vec{v}_{\rm tl}$ by relying on the second assumption that the TLF undergoes symmetric fission. Fig. \ref{lmt_vtlf} presents the velocity of the TLF (left) and the LMT (right), respectively. Again, the red and blue curves represent the central and peripheral reactions,  where the LMT distribution peaks at ${\rm LMT}=0.81$ and 0.27, respectively. 
Fig. \ref{lmt_fold_corr} presents the correlation between the LMT and the folding angle $\Theta_{\rm FF}$. It is evident that these two quantities are reversely correlated, i.e., the LMT decreases with increasing the folding angle. This correlation exists as a general feature without relying on the symmetric fission assumption, although the details of the correlations may change if this assumption is released. Nevertheless, to minimize the effect of the symmetric fission assumption, we will use the folding angle as a measurement of the reaction violence in the following discussions. 

\begin{figure}[htbp]
	\centering
	\includegraphics[width=0.48\textwidth]{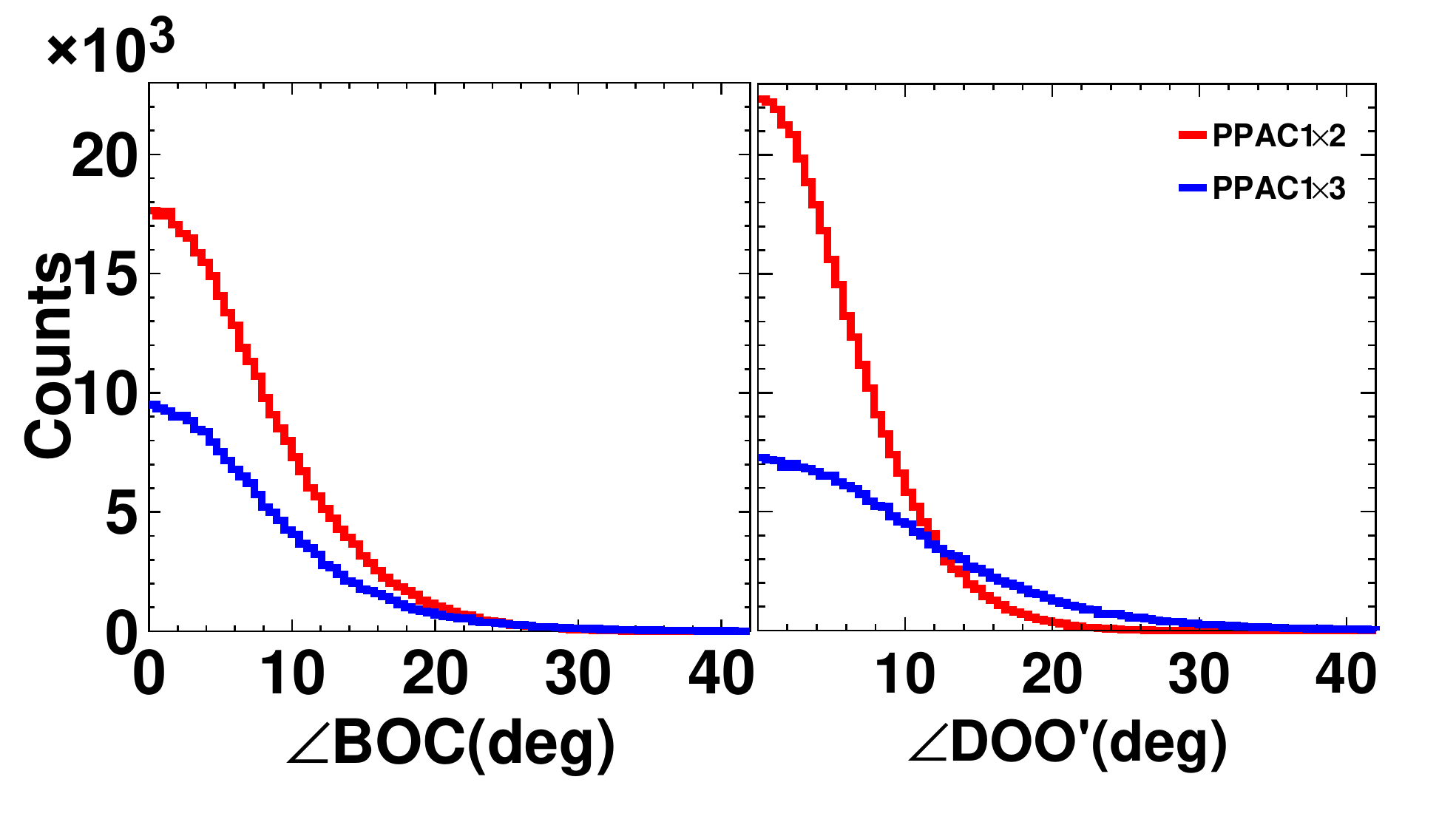}
	\caption{(Color online)  The distributions of the planarity angles $\angle BOC$ (a) and $\angle{DOO^{'}}$ (b), recalling the definition of the angles in \ref{fission_vector}.}
	\label{geo_cut}
\end{figure}

Finally, to eliminate the influence of the events located at each quantity's outer border, we impose four loose cut conditions on the experimental data. i) On the TOF distribution, $-40<\Delta TOF<40$ ns and $-50<\Delta TOF<30$ ns are applied for \pacab~ and \pacac~ events, respectively,  ii) on the folding angle, the cut of $\Theta_{\rm FF}>85^\circ$ is used to minimized the contamination from the FF correlation, and more importantly iii), the third and fourth cuts are introduced to confine the planarity of the fission events. Recalling Fig \ref{fission_vector}, it can be comprehended that the $\angle BOC$ reflects the deviation of the projection plane from the fission plane, while the $\angle{DOO^{'}}$ denotes how much the $\vec{v}_{\rm tl}$ deviates from the beam. If the fission of the rotating TLF occurs ideally in the reaction plane without being recoiled by particle emission, the two angles are zero.  Fig. \ref{geo_cut} presents the distributions of $\angle BOC$ and $\angle{DOO^{'}}$.  It is shown that both angles drop rapidly from $0^\circ$, indicating that the planarity is kept well. We set two cuts as  $\angle BOC<20^\circ$ and $\angle{DOO^{'}} <20^\circ$, which help to remove some multi-body events with the emission of undetected massive fragments causing the degradation of the planarity of the two-body fission process. Table \ref{tab:cuts} summarizes the cut conditions on the data analysis.

\begin{table}[htbp]
	\caption{The sum of value range in the experiment.}
	\begin{center}
		\begin{tabular}
			{cc}\toprule[0.65pt]   
			Physical quantity & Cut range\\
			\hline
			$\Delta TOF$ & $[-40,+40],~[-50,+30]$  \\
			$\Theta_{\rm FF}$ & $>85^\circ$ \\
			$\Delta\phi$ & $[160^\circ,200^\circ]$ \\	
			$\angle{BOC}$ & $<20^\circ$ \\	
			$\angle{DOO^{'}}$ &$<20^\circ$\\											
			\hline 
		\end{tabular}
	\end{center}
	\label{tab:cuts}
\end{table}
\vspace*{-6mm}

\section{Results and Discussions}
\subsection{Fission Distributions}

\begin{figure}[htbp]
	\centering
	\includegraphics[width=0.48\textwidth]{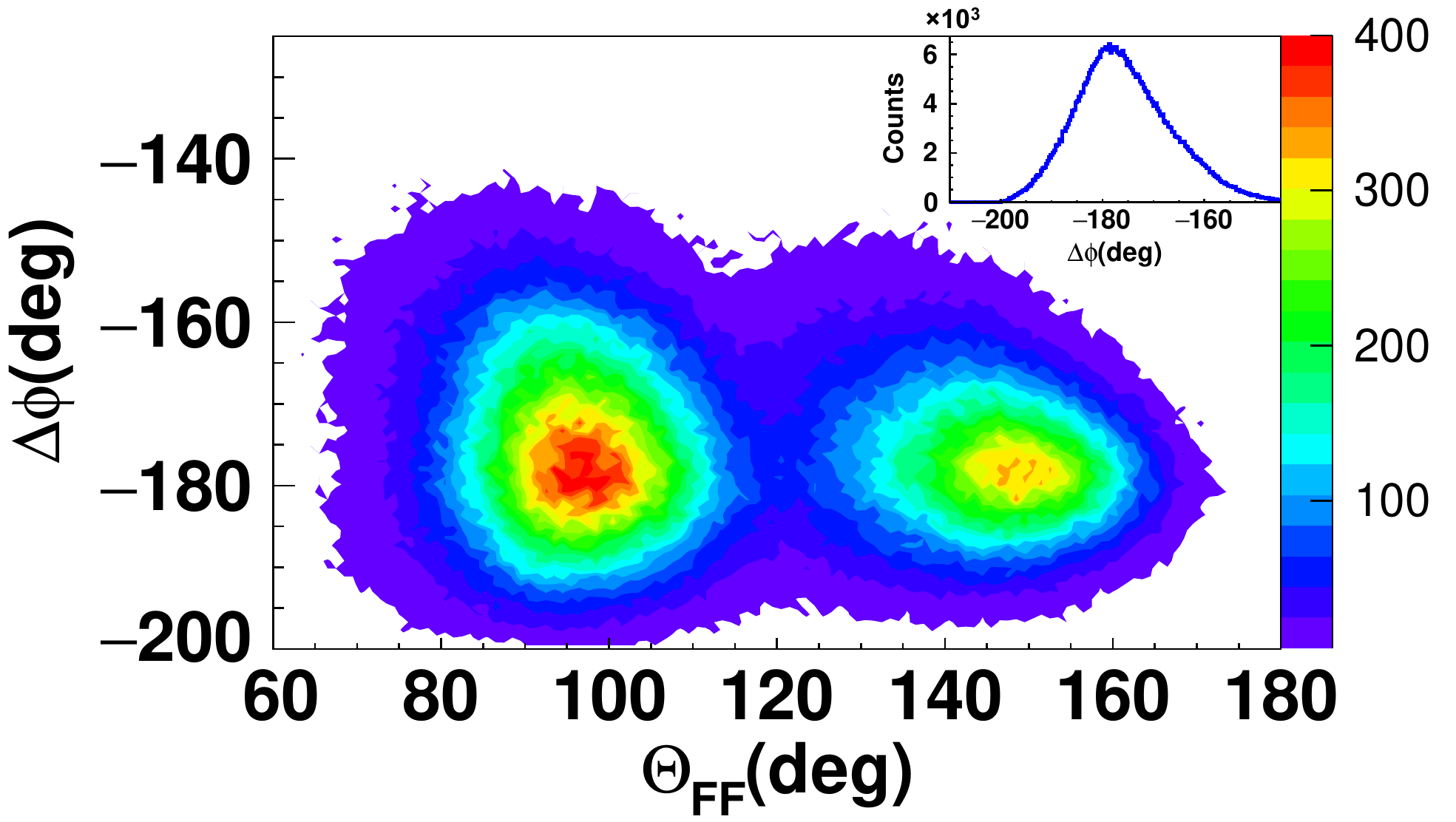}
	\caption{(Color online)  The scattering plot of the  azimuthal difference $\Delta \phi$ $vs$ the folding angle $\Theta_{\rm FF}$ in  of the FFs.}
	\label{dphi_fold}
\end{figure}

To view the fission distribution properties, we start with the azimuth correlation of the two FFs. Fig. \ref{dphi_fold} presents the correlation between the azimuthal angle difference $\Delta \phi$ and the folding angle $\Theta_{\rm FF}$. It is clearly shown that for the FFs from both the central and peripheral reactions, the most probable value situates at  $\Delta \phi=180^\circ$ following the picture that the system undergoes a  binary decay. Here we note that the  $\Delta \phi$ is a directly measurable quantity relying on no assumption. In addition, the distribution of   $\Delta \phi$  exhibits a certain broadening which depnds on the folding angle, what suggests that the emission of LCP or IMF may change the flight direction of the FF and smear the back-to-back feature of the fission event. 

\begin{figure}[htbp]
	\centering
	\includegraphics[width=0.48\textwidth]{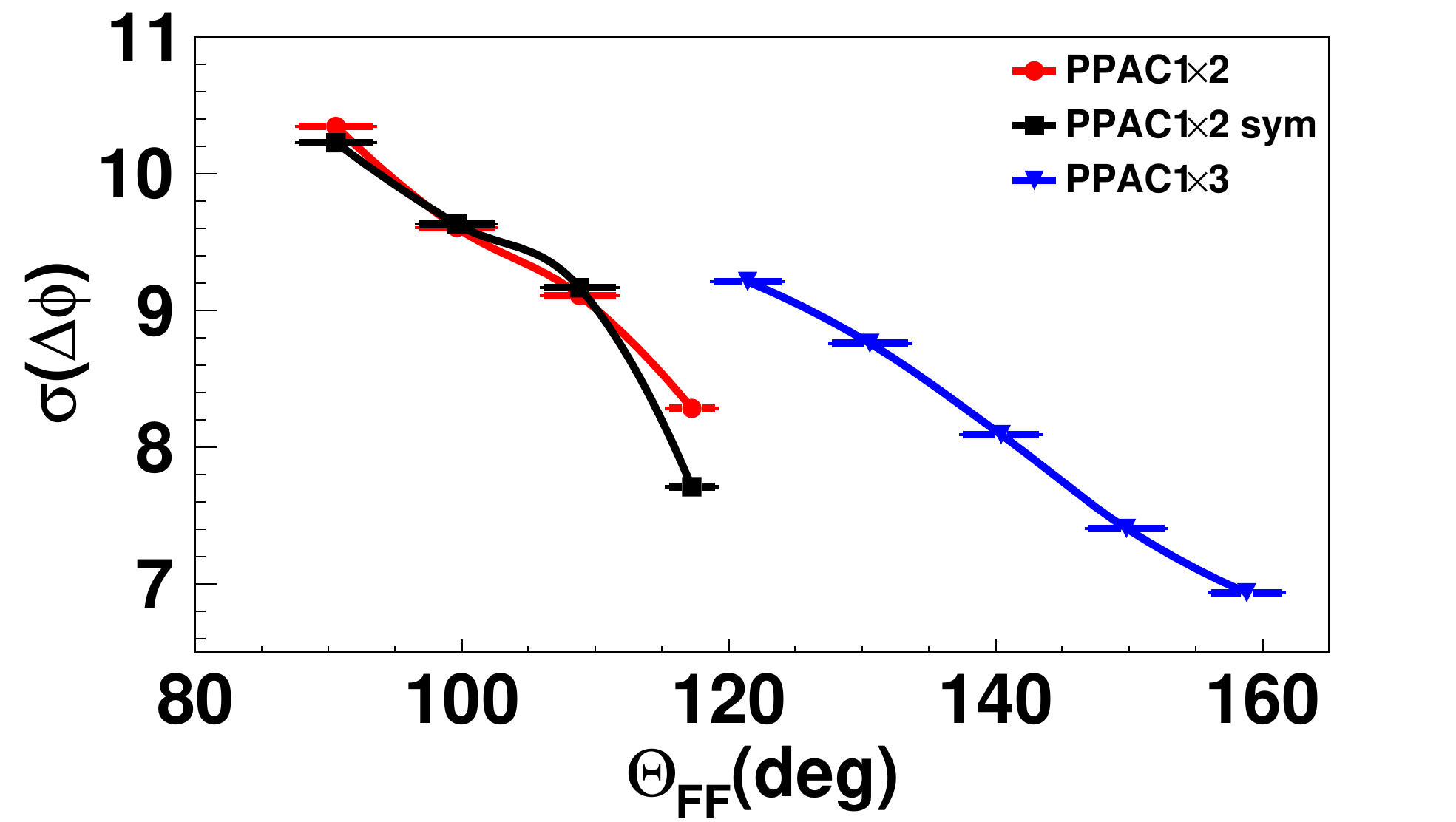}
	\caption{The standard deviation of $\sigma(\Delta\phi)$ of the azimuthal difference as a function of folding angle. The red (blue) symbols represent the \pacab~ (\pacac) events, while the black symbols represent the events in \pacab~ with an additional condition that the flight paths of the two FFs are geometrically symmetric with $\theta_1=\theta_2$.}
	\label{sig_dphi}
\end{figure}

In order to further quantify the variation of the azimuthal difference, the standard deviation $\sigma(\Delta \phi)$ of $\Delta \phi$  is extracted in each $\Theta_{\rm FF}$ bin of 10$^{\circ}$. The results are summarized in Fig. \ref{sig_dphi} for  the \pacab~ (red) and \pacac~ (blue) fission events. It is shown that  $\sigma(\Delta \phi)$ decreases with the folding angle in the whole range. To exclude the possible reason that this trend originates from the asymmetry of the geometrical locations of the PPACs, we restrict further the analysis on the events with the two FFs flying symmetrically to the beam direction, i.e., with the condition of $\theta_1=\theta_2$, where $\theta_i$ is the polar angle of the $i^{\rm th}$ fragment. The result is depicted by the black squares, where the bin width of each $\Theta_{\rm FF}$ is $\pm 3^\circ$. This condition is applicable  only for the \pacab~ fission events because these two PPACs are placed in approximate left-right symmetry with respect to the beamline. It is shown clearly that $\sigma(\Delta \phi)$ decreases with the folding angle in the same manner with the only exceptional point on the very edge of the PPAC2. It is concluded that the decreasing trend of $\sigma(\Delta \phi)$ as function of $\Theta_{\rm FF}$  is truly due to the reaction violence. Since the post-scission particle emission changes the velocity of the FF due recoil effect, the trend suggests that in the fission following the intermediate energy heavy-ion reactions, there is sufficient excitation energy left at the scission point depending on LMT. In the reactions with larger LMT, more excitation energy is left and released through the particle emission in the post-scission stage. It is consistent with the picture of fast fission, instead of statistic fission in which the excitation energy is nearly depleted at the scission point.      

Next, we investigate the other directly measurable quantity in the experiment, the TOF difference of the two FFs. In the current scheme with Eq. (2) and the current geometrical configuration that the central detector-target distances are the same for three PPACs, the TOF difference $\Delta TOF$ between the FFs originates mainly from the velocity difference  $\Delta v$ of the FFs, which indirectly reflects the mass asymmetry of the FFs. From Eq.  (2) and assuming $\Delta v/v <<1$, one writes
\begin{equation}
	\frac{\Delta v}{v}\approx \frac{v}{L}\Delta TOF +\frac{\Delta L}{L} \approx  \frac{v}{L}\Delta TOF  
\end{equation}
where the $\Delta L$ is the difference of the flight length $L$.

\begin{figure}[htbp]
	\centering
	\includegraphics[width=0.48\textwidth]{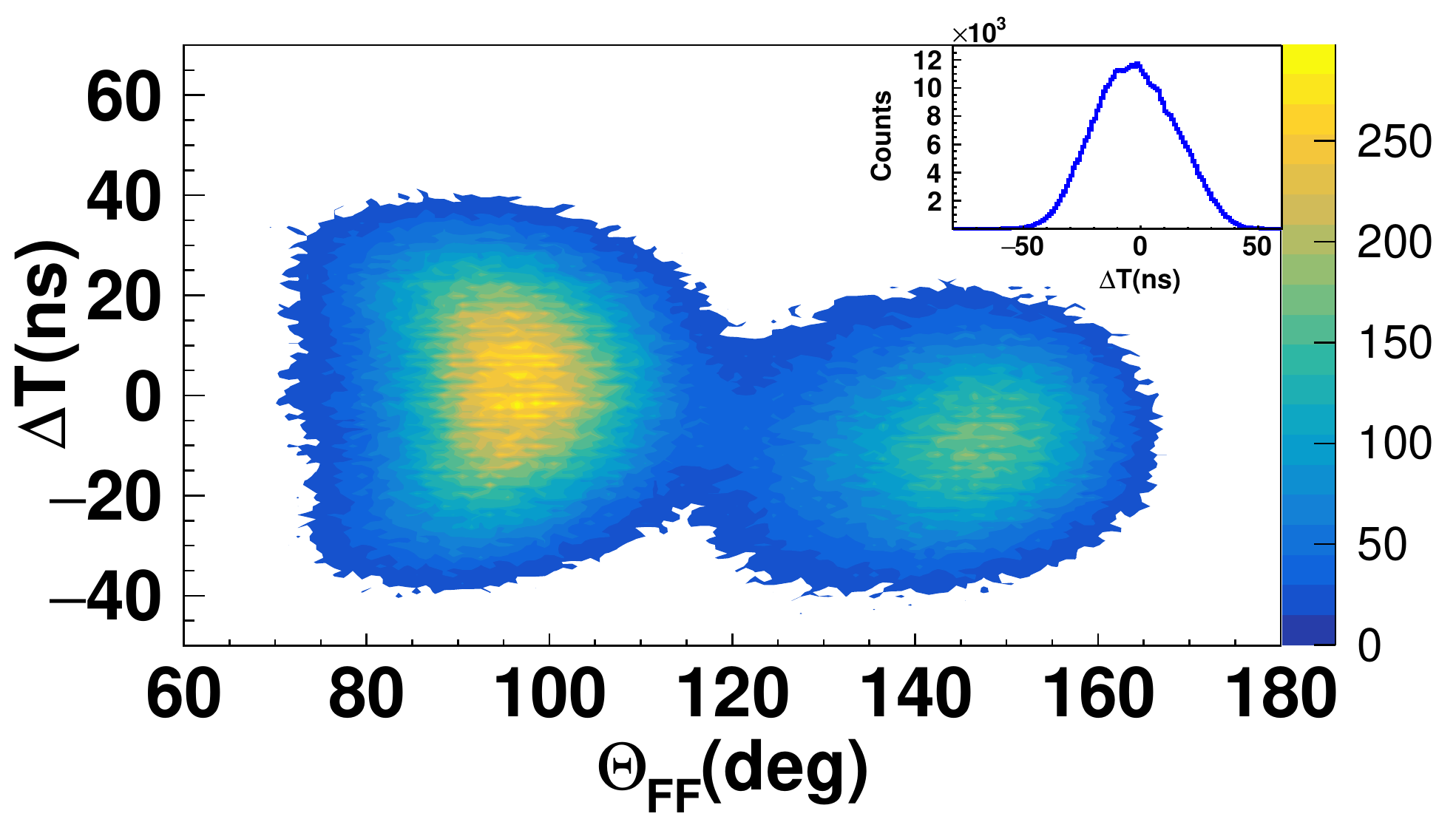}

	\caption{(Color online) The scattering plot of the TOF difference  $\Delta{TOF}$ $vs$ the folding angle $\Theta_{\rm FF}$ in two-body fission events.}
	\label{dt_fold}
\end{figure}

Fig. \ref{dt_fold} presents the correlation between $\Delta{TOF}$ and the folding angle $\Theta_{\rm FF}$. The 1-dimensional distribution of  $\Delta{TOF}$ is shown in the inset. Moving from small $\Theta_{\rm FF}$ area to large values, the broadening of $\Delta{TOF}$  decreases, showing a similar trend as $\Delta \phi$. 

\begin{figure}[htbp]
	\centering
	\includegraphics[width=0.48\textwidth]{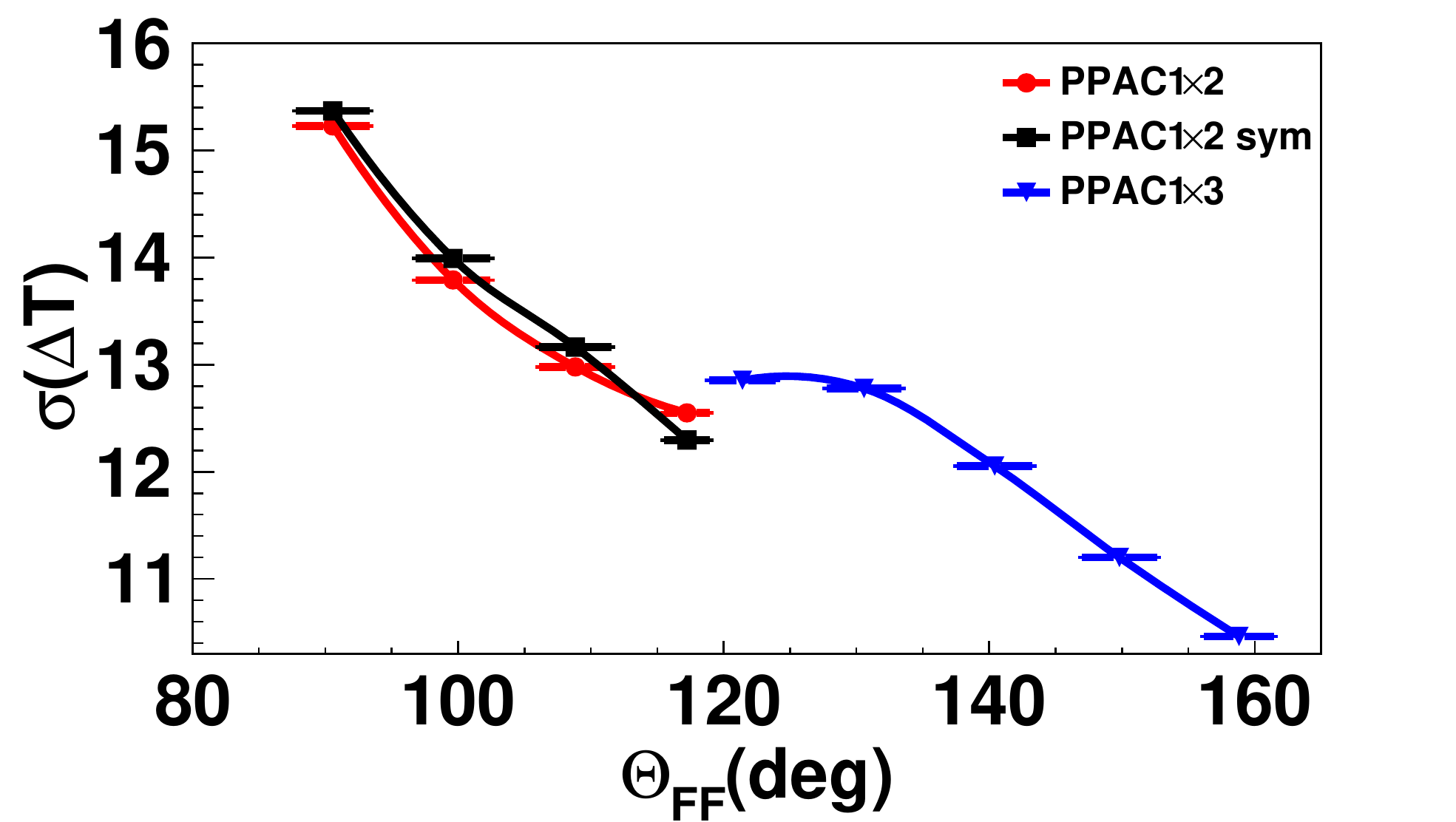}
	\caption{(Color online) The standard deviation of $\sigma(\Delta TOF)$ of the TOF difference as a function of folding angle. The red (blue) symbols represent the \pacab~ (\pacac) events, while the black symbols represent the events in \pacab~ with an additional condition that the flight paths of the two FFs are geometrically symmetric with $\theta_1=\theta_2$.}
	\label{sig_dt}
\end{figure}

Fig. \ref{sig_dt} further presents the standard deviation of the $\Delta TOF$ distribution as a function of folding angle  $\Theta_{\rm FF}$ for the \pacab~ (red) and \pacac~ (blue) fission events. Again, to minimize the possible effect of the variation of geometric conditions, we restrict the analysis to the events with symmetric geometry, i.e., $\theta_1 = \theta_2$ in the \pacab~ fission events, as depicted by the black squares in the figure. With this geometrically symmetric condition, a non-zero TOF suggests that the masses of the two fragments are not equal since the distance of the two PPACs to the target is the same. The decreasing trend of the black squares for the conditional events means that the mass asymmetry of the fission fragments decreases with increasing (decreasing) the folding angle (LMT). Moreover, in the overlapped region, the data points situate on top of each other with and without the geometrically symmetric condition, suggesting that the geometrical asymmetry is not the main reason for the variation of $\Delta TOF$. The overall descent trend of $\Delta TOF$ with $\Theta_{\rm FF}$ suggests that the mass asymmetry of the fast fission increases with the LMT of the reactions. We note that the result is independent of the assumption of Viola systematics of Eq. (3) since only the original timing information directly from the experiment is used here. 

\begin{figure}[htbp]
	\centering
	\includegraphics[width=0.48\textwidth]{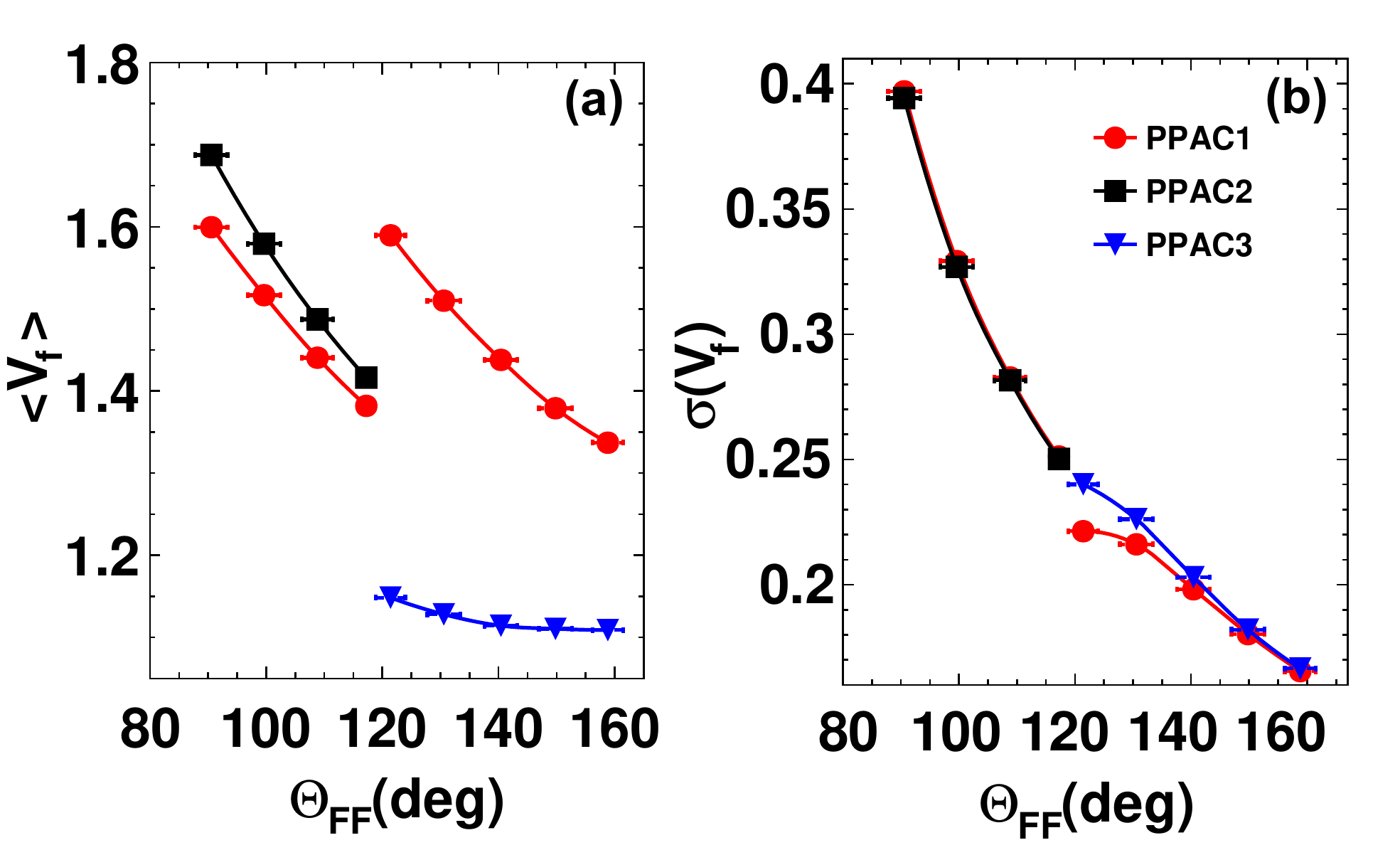}

	\caption{ (Color online) The average value of velocity distribution  (a) and its standard deviation (b) as a function of the folding angle. The blue, red, and black legends correspond to the fission fragments recorded in PPAC1, PPAC2, and PPAC3, respectively.}
	\label{v_folding}
\end{figure}

Further, we present in Fig. \ref{v_folding} the average (a) and the standard deviation (b) of the reconstructed velocities of the FFs recorded in the two PPACs as a function of folding angle. Here Eq. (1-3) are used. From panel (a), it can be seen that the average velocity value $\left<v_{\rm f}\right>$  of the fragments measured in PPAC 1 and PPAC 2 show an insignificant difference, while in PPAC3, the average velocity  $\left<v_{\rm f}\right>$ is systematically smaller in PPAC3  because this PPACs is located at a much larger angle and the velocity of the FF is counteracted mainly by the recoil TLF which is moving forward in the laboratory frame. Panel (b) presents the standard deviation of the velocity, and it is also clear that the broadening of the velocity of the FFs decreases with the folding angle. 

The distribution properties of $\sigma(\Delta \phi)$ and $\sigma(\Delta TOF)$ as a function of folding angle reflects part of the dynamic feature of fast fission that the scission point is early reached when the excitation energy of the TLF is still high. Thus the post-scission emission of particles may change the direction of the FFs. In addition, due to the incomplete de-excitation at the scission point, the statistical fluctuation enhances the variance of the masses and the velocity of the FF. This is consistent with the earlier experimental observation in ${\rm Ar+^{209}Bi}$  reactions at 25 MeV/u \cite{zjw1999}. The other important feature of fast fission, the time scale $\tau_{\rm f}$, is not measured and calls for further experimental efforts in two directions: i) By securing the first layer of the SSD telescope to measure the whole energy spectra, one can fit the spectra by including the post-scission source and derive the fission time constant $\tau_{\rm f}$ from the pre-, and post-scission particle multiplicity in statistical decay model  \cite{zjw1999-2}, ii) Alternatively, one can also secure the start timing information from the RF of the machine to measure the velocity of FFs without introducing the two mentioned assumptions. Furthermore, in this case, the rotation of the fissioning system can be reconstructed, and the  $\tau_{\rm f}$ can be derived if is given the rotational angular velocity \cite{Casini1993}.     

\subsection{Results for Light Charged Particles}
\begin{figure}[htbp]
	\centering
	\includegraphics[width=0.4\textwidth]{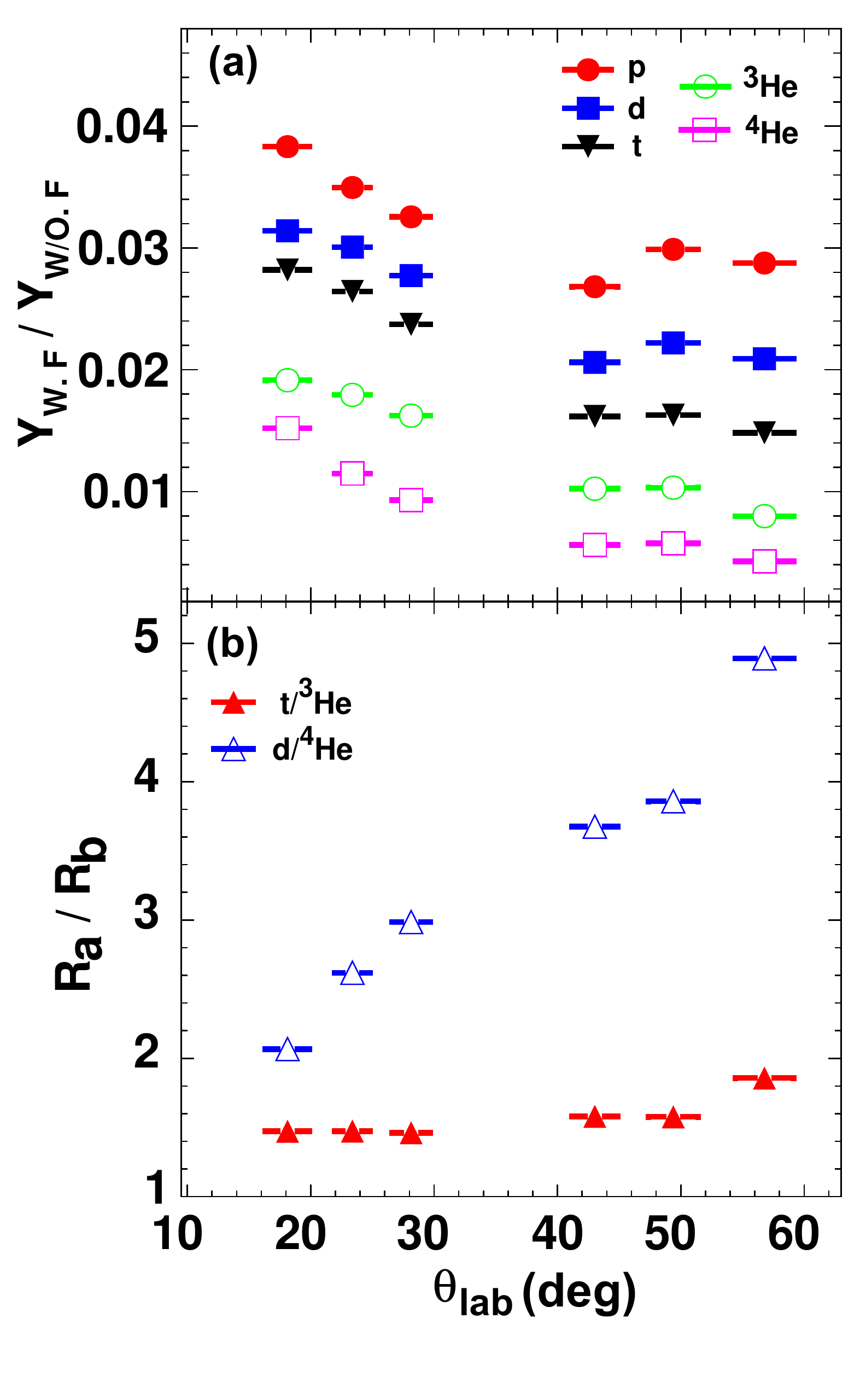}
	\caption{(a) The yield ratio of the LCPs in exclusive and inclusive events as a function of the angle in laboratory $\theta_{\rm lab}$. Panel(a) is the analysis of p, d, t, $^3$He and $^4$He. (b) shows the double ratio of  ${\rm t/^3He}$ (red) and ${\rm d/^4He}$  (blue) between the exclusive events and the inclusive events.}
	\label{lcp_angle}
\end{figure}

In this subsection, we investigate the emission of light-charged particles associated with fission. It is helpful to compare the LCP production with and without coincidence with the fission fragments to infer fast-fission particle emission characteristics. Fig. \ref{lcp_angle} (a) presents the yield ratio of various species between exclusive events (in coincidence with FFs) and inclusive events as a function of emission angle in the laboratory frame. The overall value of the ratios is in the order of a few percent since the PPACs only cover a small part of the whole space. Nevertheless,  it is shown that generally, in the whole angular range, the ratio decreases with increasing the mass of the particle. Noticing from the phase space of the LCPs in the experiment, it suggests that the dynamic emission of lighter particles is relatively easier than heavier particles in fission events. Alternatively, the dynamic emission of heavier particles, since it may happen earlier, takes away more excitation energy and reduces the probability of fast fission \cite{Pomorski2000}. 

More specifically, Fig. \ref{lcp_angle} (b) presents the double ratio of  ${\rm t/^3He}$ (red) and ${\rm d/^4He}$ (blue) between the exclusive events and the inclusive events, respectively. For the pair of ${\rm t/^3He}$, the mass of the two particles is the same while the neutron richness differs. The value of the double ratio is larger than unity, indicating that tritons are relatively easier to be emitted dynamically than ${\rm ^3He}$ in fission events. It is in qualitative agreement with the prediction of the transport model simulations \cite{Wuqh2020}, namely, the occurrence of fast fission favors the emission of neutron-rich species. For the pair of ${\rm d/^4He}$, the two particles have the same isospin content but different mass. As shown, with the increasing of $\theta_{\rm lab}$, the double ratio of {\rm d/$^4$He}, which is larger than unity again, exhibits an upward trend. It suggests similarly that the events with fission favor the production of particles with smaller masses in accordance with the results in panel (a), particularly at large angles corresponding to dynamic emissions. 

\begin{figure}[htbp]
	\centering
	\includegraphics[width=0.4\textwidth]{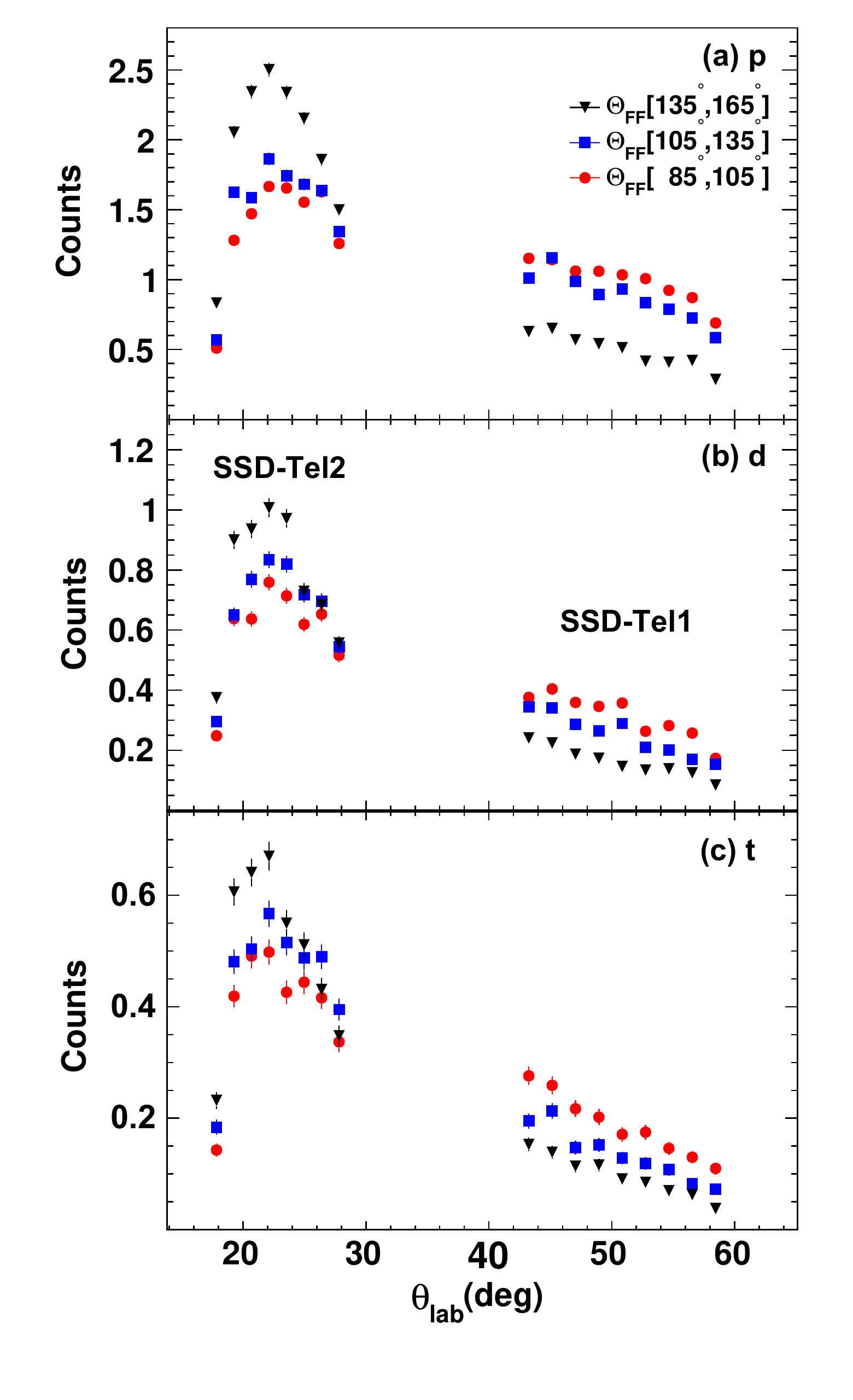}
	\caption{The angular distribution in laboratory $\theta_{\rm lab}$ for $Z=1$ isotopes in four windows of the folding angle $\Theta_{\rm FF}$ . The numbers of events in all  windows are normalized.}
	\label{theta_lcp}
\end{figure}

It is of interest to look into the angular distribution of the particle emission. Since the particle emission occurs in competition with fission, the angular distribution of the particles w.r.t. the fission axis can be used to characterize the deformation, and the angular momentum of the fissioning nuclei before scission point \cite{PBR96,SCH99}. Similarly, for the dynamic emission of the LCPs in the fast fission induced by heavy-ion reactions,  the angular distribution of the LCPs w.r.t. the fission axis carries information of the dynamic evolution of the fission process. In the current experiment, however, since the derivation of the velocity of the two FFs relies on the assumptions mentioned in Section 3, the fission axis is not precisely determined, preventing us from discussing the angular distribution w.r.t. the fission axis.  Nevertheless,  we can simply investigate the LCP emissions as a function of the laboratory angle $\theta_{\rm lab}$, which is in turn correlated with the angle of LCPs with respect to the fission plane  $\theta_{\rm fp}$  because the  PPACs cover very limited range in azimuth. Fig. \ref{theta_lcp} presents the distribution of $\theta_{\rm lab}$ for the $Z=1$ isotopes. The results of $^3{\rm He}$ and $^4{\rm He}$ are not shown because of the lower statistics. To view the effect of the reaction violence, we divide the folding angle into three windows as $85^{\circ}-105^{\circ}$, $105^{\circ}-135^{\circ}$ and $135^{\circ}-165^{\circ}$, represented by different colors, respectively. In each folding angle window, the total number of events are normalized to the same value in order to conveniently compare the features of all  $\Theta_{\rm FF}$ windows.
Interestingly, in the small angular regime with $\theta_{\rm lab} \approx 25^\circ$ covered by SSD-Tel 2, the yield of LCPs with larger folding angle (smaller LMT) is more abundant, while in the large angular regime with $\theta_{\rm lab} \approx 50^\circ$ covered by SSD-Tel 1, the opposite trend is evident. The relative difference is approximately 50\% at large angles. We note here that this difference is not due to the threshold effect because the threshold of the two SSD telescopes is very close.  It suggests that for the events with large LMT, emissions of light particles are increasingly enhanced at a large angle. It is consistent with the picture that the contribution of the intermediate velocity source increases with increasing the LMT or the violence of the collisions between the projectile and the target.

\begin{figure}[htbp]
	\centering
	\includegraphics[width=0.4\textwidth]{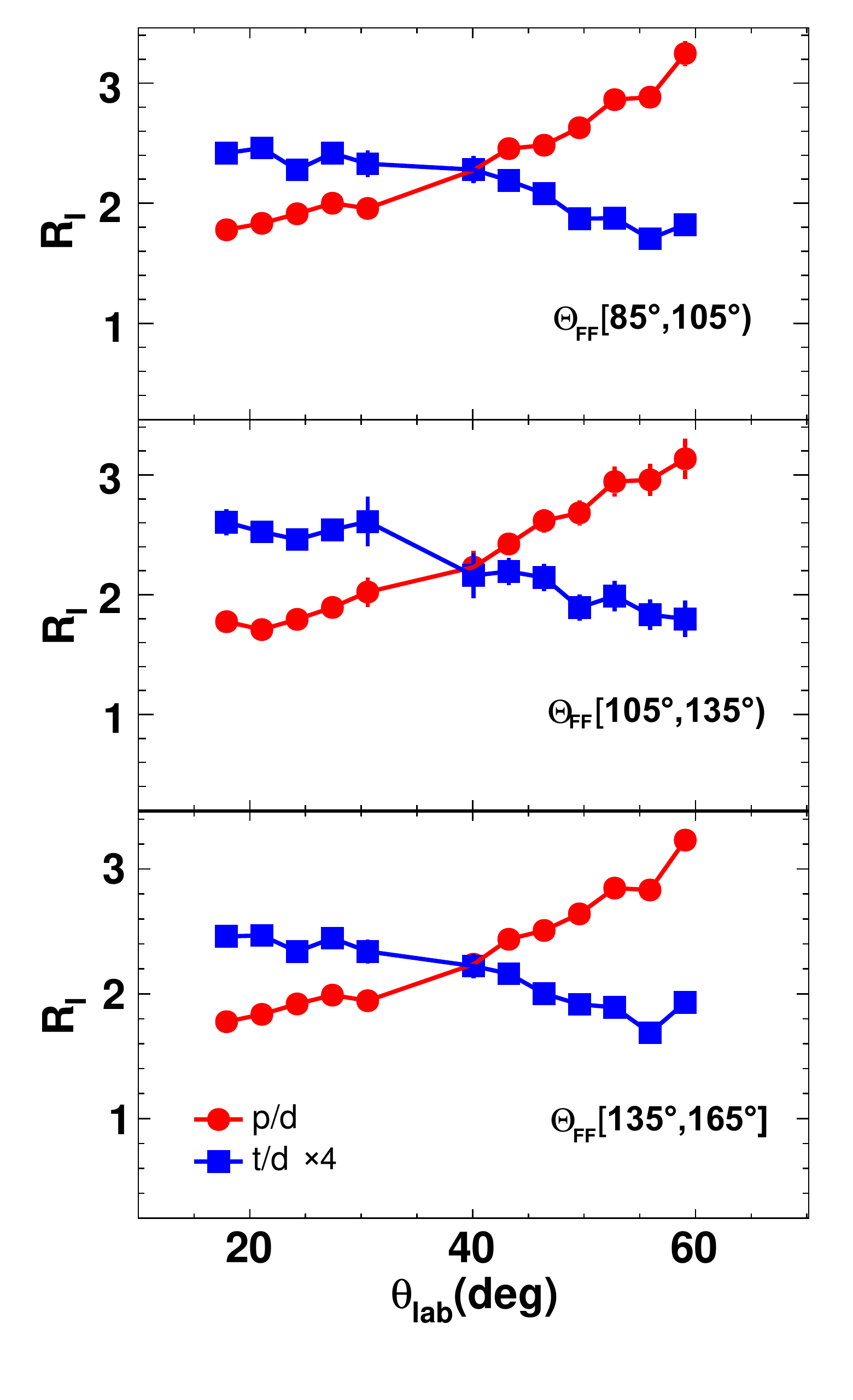}
	\caption{The yield ratio of p/d, t/d as a function of  $\theta_{\rm lab}$, respectively.}
	\label{lcp_isospin}
\end{figure}

Fig. \ref{lcp_isospin} (${\rm a-c}$) presents the isotopic yield ratio $R_I$ of p/d and t/d as a function of $\theta_{\rm lab}$  in the same three $\Theta_{\rm FF}$  windows. It is shown that in all $\Theta_{\rm FF}$  windows, the ratio of p/d increases with $\theta_{\rm lab}$ while that of t/d decreases with $\theta_{\rm lab}$. The opposite trend for these two ratios suggests that neutron-rich species tend to be emitted in the forward angle region in the fast fission event. The difference becomes increasingly enhanced if the folding angle (LMT) decreases (increases). These results are consistent with our early experimental results in the same reaction, where the inclusive production of the light-charged particles are recorded \cite{WRS2014,zy2017}. 
The angular evolution of the neutron excess of the light charged particles in coincidence with fission fragments provides a novel probe to the density behavior of nuclear symmetry energy \cite{Wuqh2020,zy2017}. Compared to the inclusive measurement, and the exclusive measurement in coincidence with fission events helps determine the event geometry so that the comparison to the transport model is better constrained. The theoretical efforts along this direction are welcomed. 

\section{Summary}

To summarize, the fission fragments and the coincident light charged particles had been measured in the experiment of $^{40}$Ar + $^{197}$Au reactions at 30 MeV/u. The fission fragments are measured using PPACs, and the LCPs are measured using SSD telescopes. Using the Viola systematics that the most probable value of the relative velocity between the fission fragments is 2.4 cm/ns, the event topology can be reconstructed, with the folding angle representing the linear momentum transfer. The broadening of the azimuthal difference $\Delta\phi$ between the two fission fragments increases with LMT, suggesting the increasing contribution of the post-scission emission, which smears the back-to-back feature of the two-body process. The difference of the time-of-flight between the two fragments also exhibits an increasing width as a function of LMT, suggesting that the mass distribution of the fission fragments tends to be broadened if more momentum is transferred to the fissioning system.

The occurrence of fast fission brings some enhancement of the emission of particles with smaller mass or higher neutron richness, as demonstrated by comparing the yield of the LCPs in the event groups with and without recording the fission fragments in coincidence. In the events followed by fast fission, the emission of LCPs becomes relatively more abundant in the intermediate angles than the forward angles when the  LMT increases, indicating more contribution of the intermediate velocity source with increasing the violence of the reaction. Commonly for all LMT windows, the contribution of the neutron-rich (neutron-deficient) particles is more abundantly emitted at a smaller (larger) angle in the laboratory frame, according to the results obtained in the previously inclusive experiment. The exclusive measurement is more beneficial than future transport model simulations aiming to extract the nuclear symmetry energy. The studies call for more attention from model simulations to reveal the isospin dynamics and constrain the nuclear symmetry energy in heavy ion-induced fission reactions at Fermi energies.     

\begin{acknowledgments}
	This work is supported by the National Natural Science Foundation of China under Grants Nos. 11961131010 and 11875174, by the Polish National Science Center under Grant No. 2018/30/Q/ST2/00185. The authors acknowledge the detector group from IMP, CAS for providing the CsI (TI) crystals and the machine staff delivering the Argon beam.
\end{acknowledgments}

\bibliography{mrcref}


\end{document}